\documentclass[%
reprint,
onecolumn,
superscriptaddress,
 amsmath,amssymb,
 aps,
prb,
]{revtex4-2}

\usepackage[utf8]{inputenc}
\usepackage[T1]{fontenc}
\usepackage{graphicx}
\usepackage{float}
\usepackage{color}
\usepackage{xr}
\usepackage{bm}
\usepackage{rotating}
\usepackage{ulem}
\usepackage{hyperref}
\usepackage{mathrsfs}

\newcommand{\DV}[1]{\textcolor{black}{#1}}
\newcommand{\DVb}[1]{\textcolor{black}{#1}}

\newcommand{\refA}[1]{\textcolor{black}{#1}}
\newcommand{\refB}[1]{\textcolor{black}{#1}}

\newcommand{\Rey}{\mathrm{Re}}
\newcommand{\Ca}{\mathrm{Ca}}

\begin{document}
	\title{Interface-induced turbulence in viscous binary fluid mixtures}
	\author{Nadia Bihari Padhan}
	\email[]{nadia@iisc.ac.in}
		\affiliation{Centre for Condensed Matter Theory, Department of Physics, Indian Institute of Science, Bangalore, 560012, India}
    \author{Dario Vincenzi}
	 \email[]{dario.vincenzi@univ-cotedazur.fr}
	\affiliation{Université Côte d’Azur, CNRS, LJAD, 06100 Nice, France}
	\author{Rahul Pandit}
		\email[]{rahul@iisc.ac.in}
	\affiliation{Centre for Condensed Matter Theory, Department of Physics, Indian Institute of Science, Bangalore, 560012, India}

\date{\today}

\begin{abstract}
We demonstrate the existence of interface-induced turbulence, an emergent nonequilibrium statistically steady state (NESS)  with spatiotemporal chaos, which is induced by interfacial fluctuations in low-Reynolds-number binary-fluid mixtures. We uncover the properties of this NESS via direct numerical simulations (DNSs) of cellular flows in the Cahn-Hilliard-Navier-Stokes (CHNS) equations for binary fluids. We show that, in this NESS, the shell-averaged energy spectrum $E(k)$ is spread over \refB{more than one decade} in the wavenumber $k$ and it exhibits a power-law region, indicative of turbulence \textit{but without a conventional inertial cascade}. To characterize the statistical properties of this turbulence, we compute, in addition to $E(k)$, the time series $e(t)$ of the kinetic energy and its power spectrum, scale-by-scale energy transfer as a function of $k$, and the energy dissipation resulting from interfacial stresses. Furthermore, we analyze the mixing properties of this low-Reynolds-number turbulence via the mean-square displacement (MSD) of Lagrangian tracer particles, for which we demonstrate diffusive behavior at long times, a hallmark of strong mixing in turbulent flows.
\end{abstract}

\maketitle


\section{Introduction}
\label{sec:intro}

Additives can lead to spatiotemporal chaos in a 
fluid, even when the inertia of the fluid is negligible and the Reynolds number $\Rey$ is low. The most notable instance of this is the phenomenon of elastic turbulence in polymer solutions~\cite{groisman2000elastic,steinberg2021elastic,datta2022perspectives}. When elastic polymers are added to a laminar Newtonian solvent, their stretching generates elastic stresses that can trigger instabilities eventually resulting in a chaotic flow, which is characterized by a power-law energy spectrum~\DV{\cite{groisman2000elastic,fouxon2007spectra,steinberg19}} and strongly intermittent fluctuations~\DV{\cite{jun2009power,singh2024intermittency}}. 
Similar chaotic regimes have been observed in low-inertia wormlike-micellar solutions~\cite{fardin2010elastic,majumdar2011universality} 
and in suspensions of microscopic rods~\cite{emmanuel2017emergence,puggioni2022enhancement,PhysRevE.110.015104} and spherical rigid particles~\cite{souzy2017stretching,turuban2021mixing}.
In contrast to conventional hydrodynamic turbulence~\DV{\cite{frisch1995turbulence}}, these examples of low-$\Rey$ turbulence 
do not rely on an energy cascade, through an inertial range, so
their main applications are in microfluidics, where additives are employed to enhance mixing~\cite{groisman01} as an alternative to 
passive or active mechanical perturbations~\cite{aref2017frontiers}. By combining theory and direct numerical simulations (DNSs) we 
uncover a new type of low-Reynolds turbulence, which is driven by interfacial fluctuations, in viscous binary-fluid mixtures. We call this \textit{interface-induced turbulence}.

A good understanding of binary-fluid mixtures is crucial for modelling emulsions~\cite{ho2022emulsion},
which have a wide variety of applications in the food~\cite{gunes2018microfluidics}, cosmetics~\cite{gilbert2013rheological}, and  pharmaceutical industries~\cite{maeki2019microfluidics, zhao2013multiphase}, often in microfluidic devices, where the enhancement of mixing is of vital importance in many \DV{situations}. In addition to its practical applications, investigations of low-\textit{Re} turbulence \DV{in systems other than viscoelastic fluids} is of fundamental interest in \DV{nonlinear} physics \DV{and} fluid dynamics. Therefore, it behooves us to explore the possibility of mixing, induced by low-inertia turbulence, in binary-fluid mixtures. We initiate such an exploration by studying a cellular flow in a two-dimensional (2D) binary-fluid system. The Cahn-Hilliard-Navier-Stokes (CHNS) partial differential equations (PDEs), which couple the fluid velocity $\bm u$ with a scalar order parameter $\phi$ that distinguishes between two coexisting phases, provide a natural theoretical framework for such flows. 
Our investigations, based on direct numerical simulations (DNSs), reveal an emergent nonequilibrium statistically steady state (NESS) with spatiotemporal chaos, which is induced by interfacial fluctuations that destabilize the laminar cellular flow. Thus, we find the elastic-turbulence analog for low-Re binary-fluid mixtures: this leads to a kinetic-energy spectrum $E(k)$, spread over several decades in the wave-number $k$, with a power-law regime that is characterised by an exponent $\simeq -4.5$. By analysing the time dependence of the total kinetic-energy $e(t)$ and its power spectrum, we characterize the transitions from the cellular flow to such turbulence, for which we demonstrate, via a scale-by-scale analysis of the kinetic energy,  that there is no significant energy cascade\DV{, and therefore the chaotic dynamics is entirely driven by the interfacial stress.} Furthermore, we elucidate how \DV{such} interfacial stress leads to global energy dissipation, even though it is responsibe for both local injection as well as dissipation of energy. Finally, we quantify the mixing properties of interface-induced turbulence by showing that the mean-square-displacement (MSD) of Lagrangian tracers displays long-time diffusive behavior \refA{with a strong enhancement of mixing with respect to the laminar regime.
}

\section{Model} 
The CHNS PDEs have been used to study multi-fluid flows, which may involve droplet interactions~\cite{scarbolo2015coalescence,pal2016binary,roccon2017viscosity,negro2023yield, elghobashi2019direct}, the evolution of antibubbles~\cite{pal2022ephemeral},
and phase separation and turbulence in such flows~\cite{perlekar2017two,shek2022spontaneous,perlekar2014spinodal,fan2016cascades}.
The two-dimensional incompressible CHNS PDEs are~\cite{pal2016binary,perlekar2017two,fan2018chns}:
\begin{flalign}
    &\partial_t \phi + \bm u \cdot \nabla \phi = M \nabla^2 \left( \frac{\delta \mathcal F}{\delta \phi}\right)\,;   \label{eq:phi}\\
    &\partial_t \omega + \bm u \cdot \nabla \omega = \nu \nabla^2 \omega + (\nabla \times \mathcal{\bm S}^{\phi})\cdot \hat e_z + f^\omega; \label{eq:omega}\\[1mm]
    &\nabla \cdot \bm u  = 0\,; \; \omega = (\nabla \times \bm u)\cdot \hat e_z\,; 
    \label{eq:incom}
\end{flalign}
$\nu$, and $M$ are the kinematic viscosity, and mobility, respectively. We write Eq.~(\ref{eq:omega}) in the vorticity-streamfunction ($\omega-\psi$) form, with
    $\bm u = \nabla \times (\psi \hat e_z)$ and $\psi = - \nabla^{-2}\omega$;
the surface stress and the Landau-Ginzburg free-energy functional are, respectively,
    \begin{eqnarray}
    \mathcal{\bm S}^{\phi} &=& -\phi \nabla \left(\frac{\delta \mathcal F}{\delta \phi}\right)\;\; {\rm{and}}\\
    \mathcal F[\phi, \nabla \phi] &=& \int_{\Omega} \left[\frac{3}{16} \frac{\sigma}{\epsilon}(\phi^2-1)^2 + \frac{3}{4} \sigma \epsilon |\nabla \phi|^2\right]d\Omega\,;\label{eq:functional}
    \end{eqnarray}
    \DVb{$\Omega$ is the spatial domain,}
    $\sigma$ is the bare surface tension, and $\epsilon$ the interfacial width.
    The first term in $\mathcal F$ is a double-well potential with minima at $\phi = \pm 1$, which correspond to two bulk phases in equilibrium; the second term is the penalty for interfaces; $\phi$ varies smoothly across an interface.

     We study the CHNS PDEs~(\ref{eq:phi})-(\ref{eq:functional}) at low $Re$, with an initially square-crystalline array of vortical structures (a \textit{cellular flow}), imposed by choosing
    \begin{equation}
        f^{\omega} = \hat e_z \cdot (\nabla \times \bm f^u) = f_0 k_f [\cos(k_f x) + \cos(k_f y)]\,,\label{eq:force}
    \end{equation}
    with amplitude $f_0$ and wave number $k_f$. Such cellular flows have been used to examine 
    the melting of this crystalline array by inertial, elastic, and elasto-inertial turbulence \DV{in viscoelastic fluids}~\cite{perlekar2010turbulence,gupta2017melting,plan2017lyapunov}. 
    For $\alpha = 0$ and $\phi(\bm r) = 0$, this system has the stationary solution
    \begin{equation}
    \omega = -\omega_0 [\cos(k_f x) + \cos(k_f y)]\,;\;\;\; \omega_0 = f_0 /\nu k_f\,.
    \label{eq:cell}
    \end{equation}
    The spatiotemporal evolution of this cellular flow depends on the Reynolds, Capillary, Cahn, non-dimensionalised friction, and P\'eclet numbers that are, respectively,
    \begin{equation}
    \mathrm{Re} = \frac{UL}{\nu}\,,
    \quad
    \mathrm{Ca} = \frac{\nu U}{\sigma}\,,\quad \mathrm{Cn} = \frac{\epsilon}{L_0}\,,
    \quad
    \mathrm{Pe} = \frac{L^2U}{M\sigma},
    \end{equation}
    with
     $U = f_0/\nu k_f^2$, $L = k_f^{-1}$, $T = \nu k_f/f_0$,
and $L_0$ the side of our square simulation domain. 
\refA{At low Re, the inertia of the mixture is negligible and, as we will show below, there is no inverse cascade of energy in the system. For this reason, we have not included a friction term in Eq.~\eqref{eq:omega}}

To characterize the mixing because of interface-induced turbulence, we introduce $N_p$ tracers into the flow. For tracer $i$ (position $\bm r^{i}_0$ at time $t_0$)
\begin{equation}
    \frac{d\bm r^{i}(t)}{dt} = \bm{v}(\bm r^i, t|\bm r^{i}_0, t_0) = \bm u(\bm r^i, t) \;, 
    \label{eq:tracer}
\end{equation}
where $\bm r^i(t)$ and $\bm v(r^i, t)$ are the position and velocity of the $i_{th}$ tracer. The mean-squared displacement (MSD) is
\begin{equation}
    \Delta r ^2 (t) = \langle |\bm r(t) - \bm r(0)|^2\rangle, 
    \label{eq:msd}
\end{equation}
where $\langle \cdot \rangle$ denotes the average over the $N_p$ particle trajectories.



\begin{figure}
{\includegraphics[width=0.7\textwidth]{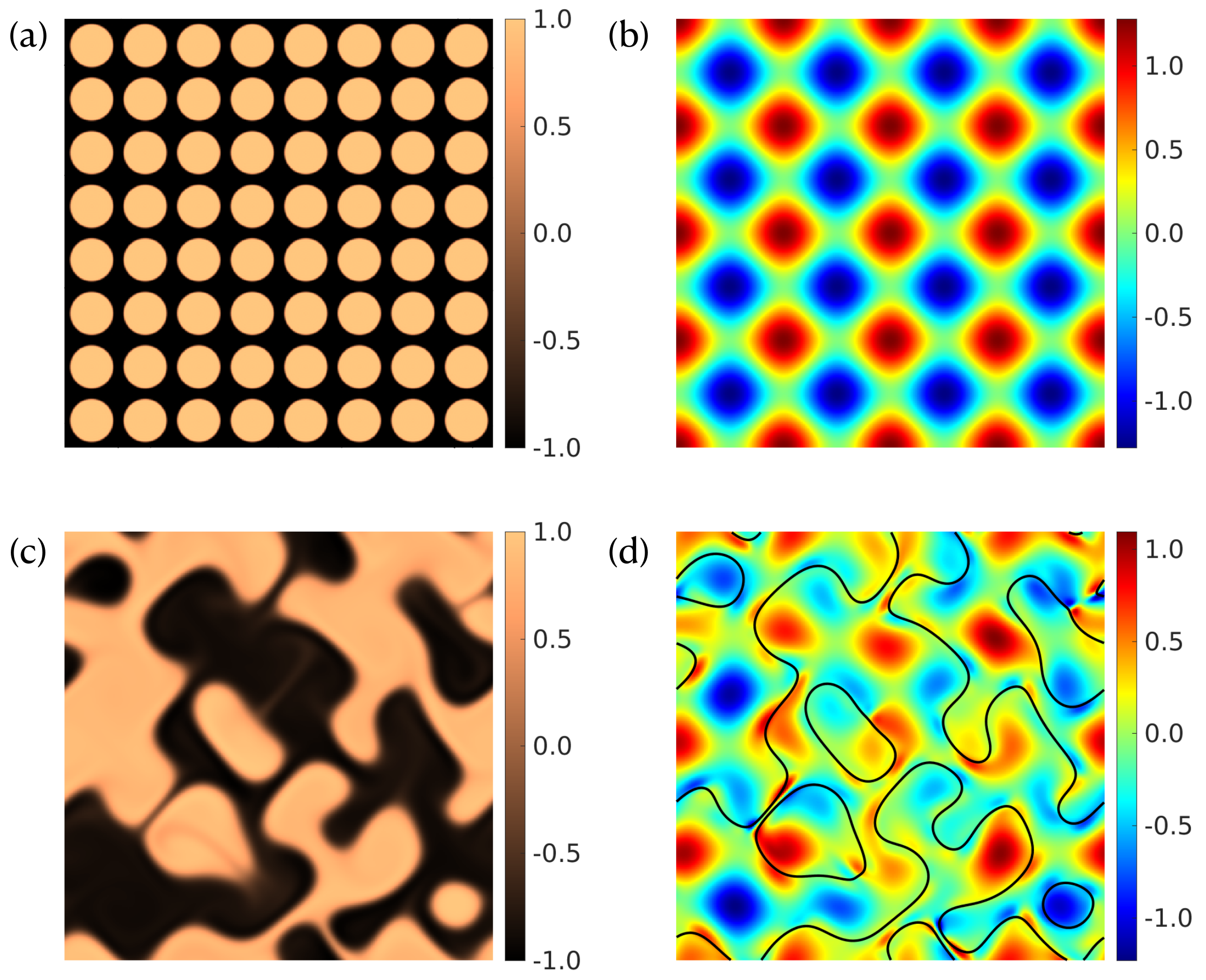}
}
\caption{\label{fig:pcolor} Pseudocolor plots: (a) the initial condition for the $\phi$ field [Eq.~(\ref{eq:init})]; (b) steady-state laminar solution [Eq.~(\ref{eq:omega})] for the vorticity $\omega$, with no droplets or friction; (c) illustrative $\phi$ field in the chaotic regime (Capillary number $\mathrm{Ca} = 0.15$); and (d) $\omega$, corresponding to subplot (c), and with the overlaid $\phi = 0$ contour lines (in black). \DV{See the Supplemental Material~\cite{supmat} for the corresponding movie.}}
\end{figure}

\section{Numerical methods and initial conditions}
\label{subsec:DNS}
We carry out pseudospectral DNSs (parameters in Table I in the Supplemental Material~\cite{supmat}) of the CHNS PDEs~(\ref{eq:phi})-(\ref{eq:functional}), with periodic boundary conditions, a square $(2\pi \times 2\pi)$ box, $512^2$ collocation points~\cite{canuto2012spectral,pal2016binary,padhan2023activity,padhan2023unveiling}, the $1/2$-dealiasing scheme, and a semi-implicit exponential time difference Runge-Kutta-2 method~\cite{cox2002exponential} for time integration. To resolve interfaces, we have three computational grid points in interfacial regions.  We obtain $\bm v$ from $\bm u$ via bilinear interpolation at off-grid points and a first-order Euler scheme for Eq.~(\ref{eq:tracer})~\cite{benzi2010inertial,verma2020first}. The initial condition [Fig.~\ref{fig:pcolor}(a)] comprises $N_d$ circular droplets~\footnote{We have checked explicitly that our results are independent of the initial arrangements and sizes of the droplets.}; droplet $i$, centered at $(x_i, y_i)$, has radius $R_i$: 
    \begin{equation}
    \phi(x, y, t=0) =\sum_{i=1}^{N_d} \tanh\left[\epsilon^{-1}\left(R_i - \sqrt{(x-x_i)^2 + (y-y_i)^2}\right)\right] \,,
        \quad
    \omega(x,y, t=0) = 0 \,. \label{eq:init}
    \end{equation}
In Fig.~\ref{fig:pcolor}(b) we show a pseudocolor plot of $\omega$ for the cellular solution~(\ref{eq:cell}),
for the single-fluid case \refA{($\phi=0$)}. 

\section{Results}
\label{sec:results}

We consider $\Rey = 1 < \Rey_c = \sqrt{2}$, 
the single-fluid \refA{($\phi=0$)} critical Reynolds number, given the cellular forcing we use~\cite{gotoh1984instability}. We choose $\Rey < \Rey_c$ to exclude inertial instabilities, so that we can focus only on interface-induced dynamics.
Our DNSs reveal that the second phase leads to interfaces whose fluctuations can destabilise this cellular flow and yield \textit{interface-induced turbulence}, a NESS with spatiotemporal chaos.
In Figs.~\ref{fig:pcolor}(c) and (d) we present pseudocolor plots, of $\phi$ and $\omega$, respectively, for $\Ca = 0.15$, which illustrate the breakdown of the cellular flow in Fig.~\ref{fig:pcolor}(b) \DV{(see also the corresponding movie in the Supplemental Material~\cite{supmat})}. 
\DV{Moreover, the time series of the rescaled total energy $e(t)/e_0$, with $e_0=U^2$, shows that, as $\Ca$ is varied, the system undergoes a \DVb{non-monotonic} sequence of transitions between periodic regimes and spatiotemporally chaotic NESSs at low $Re$ (see Fig.~\ref{fig:en_time}).
In the Supplemental Material~\cite{supmat},}
we examine the above cellular-to-spatiotemporally chaotic transitions via \DV{additional}  plots of \DV{the time series of the total energy $e(t)$, its frequency power spectrum,
and pseudocolor plots of the vorticity and the energy spectrum for a wide range of Ca.} 
\refA{It is interesting to note that a non-monotonic sequence of periodic and chaotic states is also observed in low-Re viscoelastic fluids as the fluid elasticity is varied \cite{gupta2017melting}.}
\begin{figure}
\vspace{-2cm}
\centering
{\includegraphics[width=\textwidth]{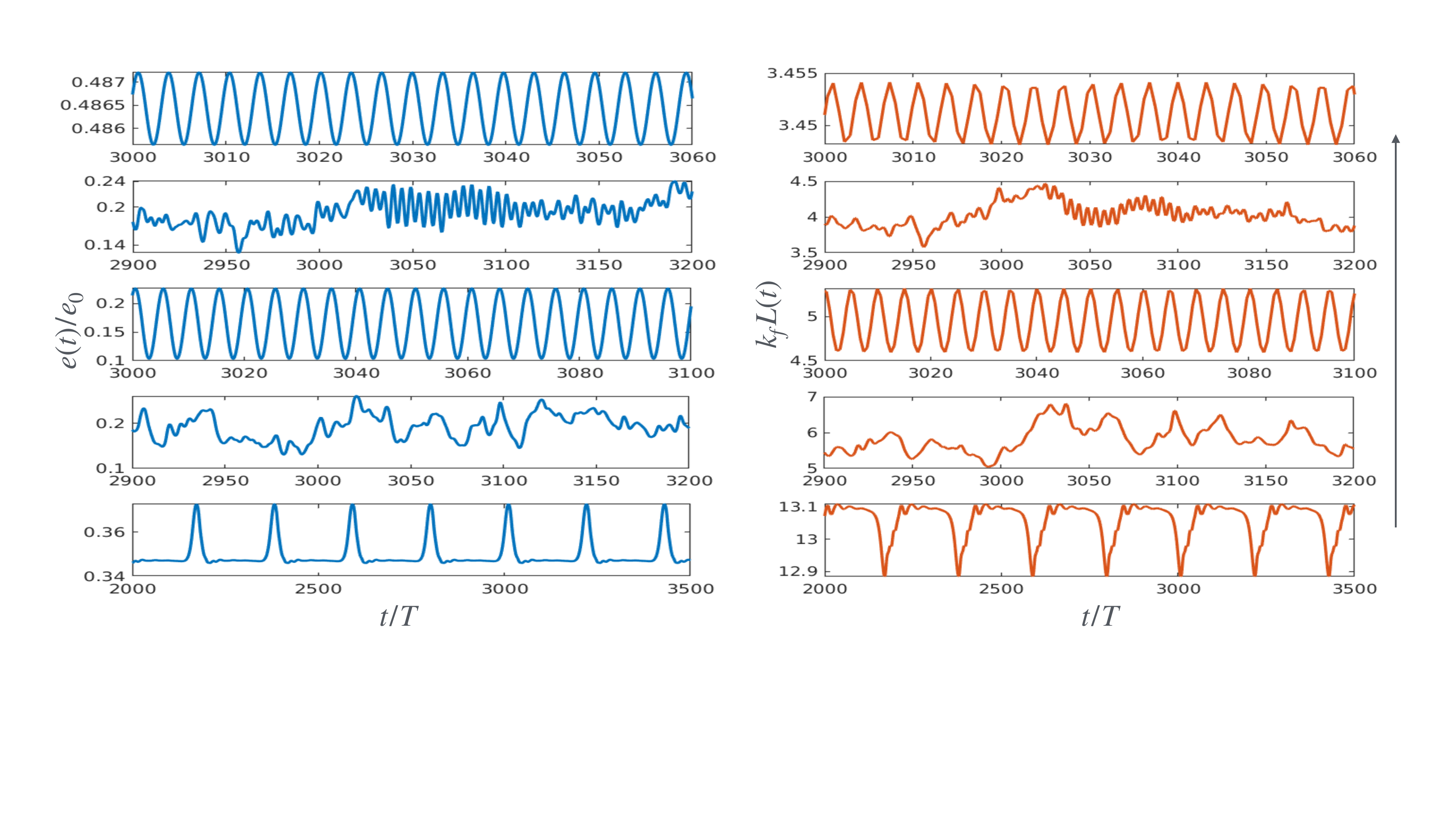}
\put(-520, 200){\rm {\bf(a)}}
\put(-255, 200){\rm {\bf(b)}}
\put(-1,100){\large\rm {\begin{turn}{90}$\Ca{\hspace*{6cm}}$\end{turn}}}
}
\caption{\label{fig:en_time} Plots of (a) $e(t)/e_0$ \refB{and (b) $k_f L(t)$} versus the rescaled time $t/T$: from bottom to top $\mathrm{Ca}=0.1,0.15,0.16,0.18,0.6$. For $\mathrm{Ca}=0.1,\,0.16,$ and $\mathrm{Ca}=0.6$, the state shows periodic oscillations in $t$; by contrast, the state is temporally chaotic at $\mathrm{Ca}=0.15$ and $\mathrm{Ca}=0.18$.
}
\end{figure}
\begin{figure*}
\includegraphics[width=\textwidth]{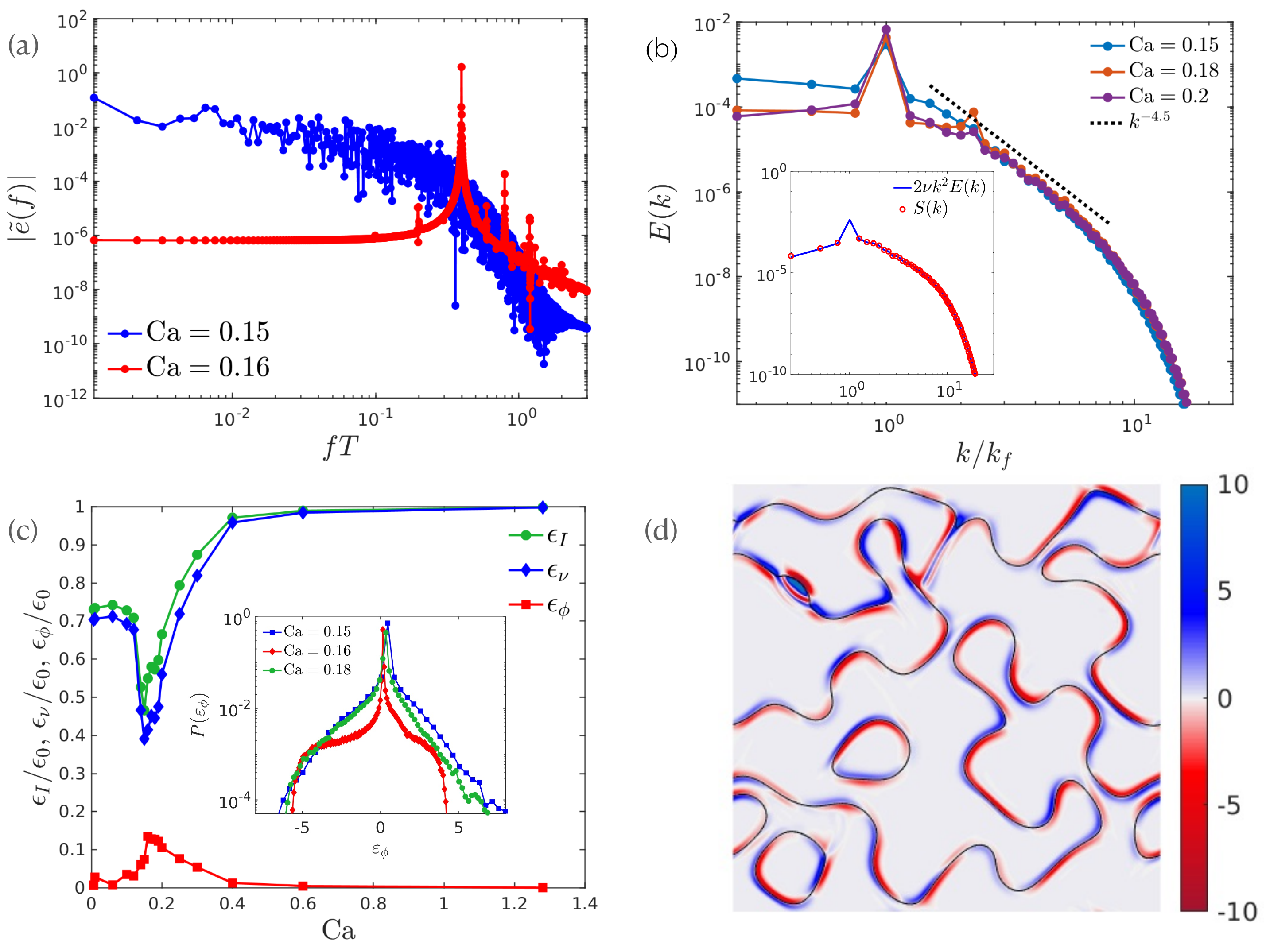}
\caption{\label{fig:fig_spec} Log-log plots of (a) the frequency power spectrum $|\tilde{e}(f)|$ versus the scaled frequency $fT$ for $\Ca = 0.15$ and $\Ca = 0.16$ and (b) the averaged energy spectrum $E(k)$ in the spatiotemporally chaotic NESS for \refB{$\Ca = 0.15, 0.18, 0.2$}; inset: the $k$-dependence of the viscous contribution $2\nu k^2 E(k)$ and the elastic-transfer term $S(k)$ for $\Ca = 0.15$; the black line suggests $E(k) \sim k^{-4.5}$. (c) Plots versus $\mathrm{Ca}$ of $\epsilon_{I}$, $\epsilon_{\nu}$, and $\epsilon_{\phi}$, the contributions of different dissipation terms in Eq.~(\ref{eq:enbudget}); inset: the PDF of the \textit{local} interfacial-stress contribution $\varepsilon_{\phi}$, for different values of $\mathrm{Ca}$. (d) The pseudocolor plot of $\varepsilon_{\phi}$ for $\mathrm{Ca} = 0.15$ at a representative time.}
\end{figure*}
\begin{figure*}
{\includegraphics[width=\textwidth]{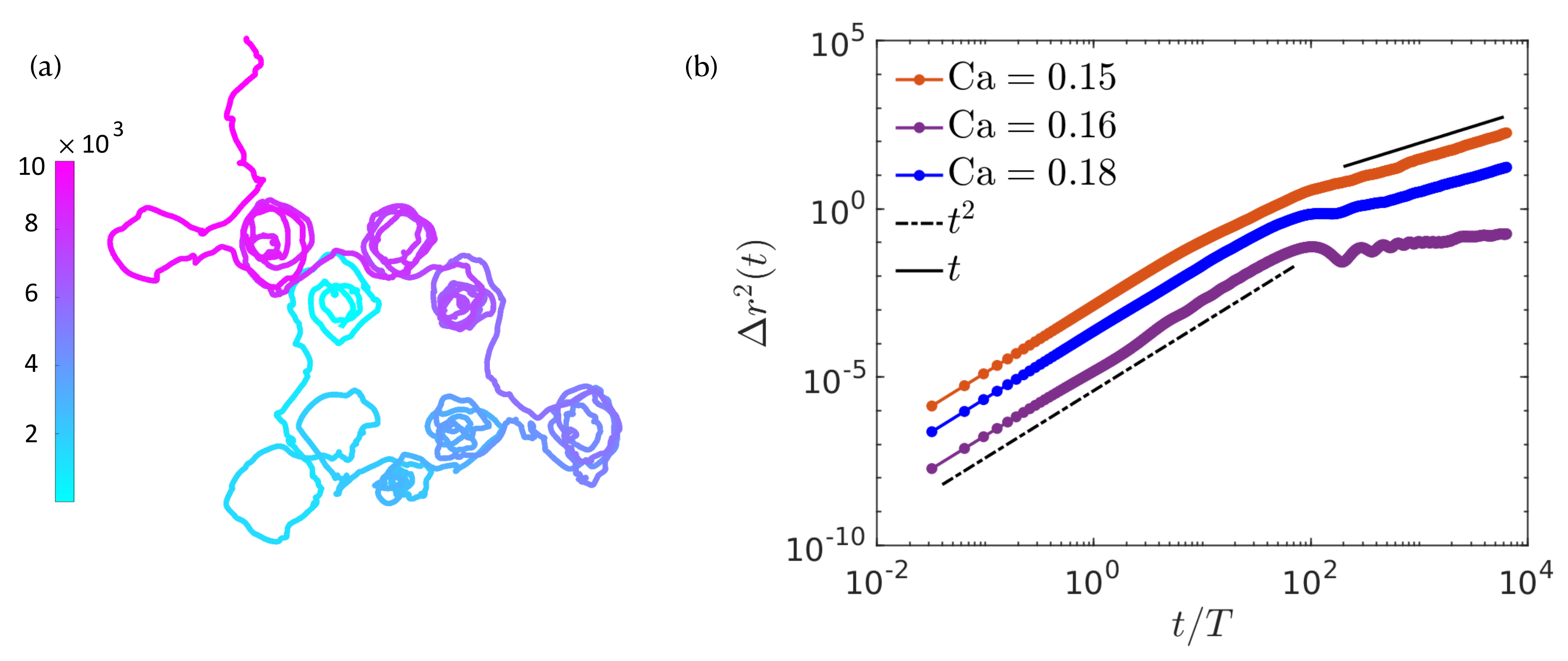}
}
\caption{\label{fig:msd_traj}(a) Illustrative plot of a tracer trajectory for $\Ca = 0.15$; the colorbar shows the simulation time. (b) Log-log plots of the 
mean-square displacement (MSD) (averaged over $N_p= 1024 \times 128$ particles), showing short-time   ballistic ($\sim t^2$) and long-time diffusive ($\sim t$) asymptotes for the spatiotemporally chaotic NESSs ($\Ca = 0.15$ and $\Ca = 0.18$); only ballistic behavior appears for the periodic time evolution at $\Ca=0.16$.}
\end{figure*}

\refB{In Fig.~\ref{fig:en_time}(b), we also show the time series of the coarsening length $L(t) = 2\pi\sum_k \mathscr{S}(k, t)/\sum_k k \mathscr{S}(k, t)$, where $\mathscr{S}(k, t)$$ = \sum_{k\leq k'< k+1} |\hat\phi_k(k', t)|^2$. Coarsening is arrested in all the regimes, and the behavior of the coarsening length, either periodic or chaotic, parallels that of the energy. We shall see below that the arrest of coarsening favors the emergence of the chaotic regime, because it ensures a sufficiently large interface between the two fluids. However, this alone is not sufficient to generate a chaotic regime: the interface must be deformable enough to create large stresses between the two fluids, which is clearly not the case at large Ca.}

We turn now to spatiotemporal properties. In Fig.~\ref{fig:fig_spec}(a) we give log-log plots of the power-spectrum \DV{of the total energy,} $|\tilde{e}(f)|$, versus the normalized frequency $fT$. For $\Ca = 0.16$, this spectrum shows a single dominant peak, a signature of temporal periodicity;  by contrast, for $\Ca = 0.15$, we see a broad power-spectrum, which indicates that $e(t)$ is chaotic.
In Fig.~\ref{fig:fig_spec}(b) we characterise the spatial distribution of the kinetic energy via a log-log plot of the shell-averaged energy spectrum $E(k)$ versus the wave-number $k$ (see \cite{supmat} for the definition), in the spatiotemporally chaotic NESS \refB{for $\Ca = 0.15,0.18,0.2$}. Over a small range of $k$, $E(k) \sim k^{-4.5}$
[black line in Fig.~\ref{fig:fig_spec}(b)]; this power-law exponent \refA{indicates that, in the regime of interface-induced turbulence, the flow is large-scale and smooth.}
\refA{Note that a} spectrum steeper than $k^{-3}$ is also a characteristic feature of elastic turbulence in polymer solutions \cite{fouxon2007spectra,steinberg2021elastic}.
\refB{By analogy with elastic turbulence, where the slope of the energy spectrum has been found to vary between $-4.6$ and $-3$ depending on the flow configuration,
we expect the slope of the energy spectrum in interface-induced turbulence not to be universal, but to depend on the specific forcing and boundary conditions.}

Unlike inertial fluid turbulence, the low-$Re$ interface-induced turbulence we consider does not show an
energy cascade. We demonstrate this via the following scale-by-scale kinetic-energy-budget equation~\cite{perlekar2019kinetic,alexakis2018cascades,verma2019energy}:
\begin{equation}
\partial_t E(k, t)=T(k,t)-2 \nu k^2 E(k,t)+S(k,t)+F(k,t)\,,\label{eq:en_budget}
\end{equation}
\DV{where $S(k, t)$ is \DVb{the contribution of the interfacial stress}, $T(k,t)$ is the nonlinear energy transfer, and $F(k,t)$ the energy-injection term \DV{(see \cite{supmat} for the definitions)}.}
In the inset of Fig.~\ref{fig:fig_spec}(b), we present the $k$-dependence of the viscous contribution $2\nu k^2 E(k)$, in blue, and, in red, the  contribution of the interfacial stress, $S(k)$, \refA{in the statistically stationary state}; both these terms are equal for all $k$, except at the forcing wave-number $k_f = 4$. 
\refA{Equation~\eqref{eq:en_budget} therefore implies that $T(k)$ is negligible at all $k\neq k_f$. We have also calculated the ratios  $\vert T(k)/S(k)\vert$ and $\vert T(k)/2\nu k^2 E(k)\vert$ vs $k$, and we have indeed found that the transfer term is at least one order of magnitude smaller than the dissipation and interfacial-stress terms for all $k$.}
In fluid turbulence, inertia plays a pivotal role in transferring energy from the energy-injection wavenumber(s) to other wavenumbers, and $T(k)$ is non-zero for most $k$. By contrast, in the interface-induced turbulence we consider, inertia is negligible, and energy in wavenumbers other than the injection wavenumber is solely attributable to $S(k)$, which is balanced by $2\nu k^2E(k)$; hence, $T(k)$ is negligibly small in Eq.~(\ref{eq:en_budget}). This energy transfer by interfacial stresses is a unique property of low-$Re$ interface-induced turbulence and distinguishes it clearly from fluid turbulence.
\DV{It is also useful to study} the energy-budget equation
\begin{equation}
    \frac{de(t)}{dt} = \epsilon_I - \epsilon_{\nu} - \epsilon_{\phi}\,; \label{eq:enbudget}
\end{equation}
$\epsilon_I = \langle \bm f^{u}\cdot \bm u\rangle_{\bm x}$ is the mean energy-injection rate, $\epsilon_{\nu} = -\langle \bm u \cdot \nu \nabla^2 \bm u\rangle_{\bm x}$ is the mean energy-dissipation (viscous) rate, $\epsilon_{\phi} = -\langle \varepsilon_{\phi}\rangle_{\bm x} = -\langle \bm u \cdot \mathcal{\bm S}^{\phi}\rangle_{\bm x}$ is the additional mean dissipation because of interfaces, and $\langle\cdot\rangle_{\bm x}$ denotes the space average.
We plot 
$\epsilon_{I}$, $\epsilon_{\nu}$, and $\epsilon_{\phi}$ versus $\Ca$ in Fig.~\ref{fig:fig_spec}(c).
At intermediate values of $\Ca$, $\epsilon_{\phi} > 0$; i.e., \textit{globally}, the interfacial 
contribution to the energy budget is dissipative. However, the interfacial stress both injects and dissipates energy \textit{locally}, as we demonstrate by plotting, in \DV{the inset of} Fig.~\ref{fig:fig_spec}\DV{(c)}, the probability distribution functions (PDFs) of $\varepsilon_{\phi}$, the local dissipation because of interfaces. The fat tails of this PDF exhibit that $\varepsilon_{\phi}$ shows large fluctuations that are both positive and negative.
\DV{The pseudocolor plot of $\varepsilon_{\phi}$ for $\Ca = 0.15$ in
Fig.~\ref{fig:fig_spec}(d) also confirms that $\varepsilon_\phi$ is concentrated at the interface between the two fluids.}
Therefore, the turbulent behavior, which we uncover by the energy-budget analysis~(\ref{eq:enbudget}), is attributable solely to the presence of interfaces in the flow, \DV{and is observed at intermediate values of Ca}. For low values of $\Ca$ (large $\sigma$), $\epsilon_{\phi}$ is low because the interfaces are so energetic that their energy surpasses the kinetic energy of the flow: thus, droplets coalesce, interfaces do not break-up, and the interfacial length is minimal. For high values of $\Ca$ (low $\sigma$), the interfacial energy is so low that it hardly affects the flow, and the system retains the cellular structure of the applied force; and the energy injection and viscous dissipation balance, i.e., $\epsilon_I/\epsilon_0 = \epsilon_{\nu}/\epsilon_0 = 1$, and $\epsilon_{\phi}/\epsilon_0 = 0$, with $\epsilon_0 \equiv f_0 U$.
\refB{The inset of 
Fig.~\ref{fig:fig_spec}(c) also shows a substantial difference in the statistics of $\varepsilon_\phi$ 
for the periodic ($\mathrm{Ca}=0.16$) 
and the chaotic ($\mathrm{Ca}=0.15, 0.18$) cases.
Indeed, the $\mathrm{Ca}=0.16$ case is dominated by events with very small $\varepsilon_\phi$, and the PDF of $\varepsilon_\phi$ rapidly drops as $\vert\varepsilon_\phi\vert$ deviates from zero. In contrast, the probability of moderate fluctuations of $\varepsilon_\phi$ remains significant in the chaotic cases.}

One of the intriguing properties of interface-induced turbulence is that it enhances mixing even at low $Re$, \DV{which makes this phenomenon of great interest for microfluidic applications}. We quantify such mixing properties by investigating the dispersion of tracer particles in the flow
[Eqs.~(\ref{eq:tracer}) and
(\ref{eq:msd})]. 
In Fig.~\ref{fig:msd_traj}(a), we depict a representative tracer trajectory in the spatiotemporally chaotic NESS for $Ca = 0.15$; the colorbar shows the simulation time. 
Initially, the particles get trapped within vortices, but, when an interface moves through these vortices, it facilitates particle transfer to other vortices. 
We plot the MSD [Eq.~(\ref{eq:msd})] versus $t$ for $\Ca=0.15$, $\Ca=0.16$, and $\Ca=0.18$ in Fig.~\ref{fig:msd_traj}(b).  
For the chaotic NESSs ($\Ca=0.15$ and $\Ca=0.18$) the small- and large-$t$ asymptotes of the MSD can be fit to the power-law-form $ \langle r^2(t) \rangle \sim t^{\beta}$, with short-time ballistic behavior $\beta = 2$, and long-time diffusive behavior $\beta = 1$, because of strong mixing via interface-induced turbulence. 
If the state is periodic, e.g., for $\Ca = 0.16$, the MSD shows only ballistic behavior and then trapping into a vortical cell at longer times. 

\section{Conclusions}
\label{sec:conclusions}
We have demonstrated how interfaces in a binary-fluid mixture can disrupt low-$Re$ cellular flows by precipitating instabilities that lead to interface-induced turbulence, the binary-fluid analog of elastic turbulence in fluids with polymer additives~\cite{groisman2000elastic,steinberg2021elastic,datta2022perspectives}. We have explored the transitions from cellular flows, to flows with spatiotemporal crystals, and, eventually, to a NESS with interface-induced turbulence. We have characterised these states via the energy time series $e(t)$, its frequency power spectrum $|\tilde{e}(f)|$, the energy spectrum $E(k)$, the energy budget [Eqs.~(\ref{eq:en_budget}) and
\ref{eq:enbudget}], and the MSD of Lagrangian tracers [Eqs.~(\ref{eq:tracer}) and (\ref{eq:msd})]. The low-$\Rey$ interface-induced turbulence that we have uncovered exhibits the following distinctive properties: (a) $|\tilde{e}(f)|$ is significant over a broad range of frequencies $f$; (b) a power-law regime with $E(k) \sim k^{-4.5}$, with a power that is different from its counterpart in 2D fluid turbulence with no friction~\cite{boffetta2012two,pandit2017overview}; (c) a scale-by-scale energy transfer [Eq.~(\ref{eq:en_budget})] with negligible inertial contribution $T(k)$; (d) an MSD of tracers that crosses over from ballistic to diffusive behaviors, indicating strong mixing. \DVb{Cellular flows have been used in experimental studies of elastic turbulence 
\cite{liu2012oscillations};}
we therefore look forward to experimental confirmations of our predictions for low-$\Rey$ interface-induced turbulence in such flows. \refA{In this regard, we note that in our study the two fluids have same kinematic viscosity. Nevertheless, the turbulent state that we have identified is induced solely by the stresses at the interface between the two fluids. We therefore expect it to persist when the kinematic viscosities of the two fluids differ, even though the range of parameters over which interface-induced turbulence is observed is likely to depend on the viscosity ratio.
}
\begin{acknowledgements}
We thank K.V. Kiran \DVb{and S.~Mukherjee} for valuable discussions. NBP, DV, and RP thank the Indo-French Centre for Applied Mathematics (IFCAM) for support
\DV{and the Isaac Newton Institute for Mathematical Sciences for support and hospitality during the programme ‘Anti-diffusive dynamics: from sub-cellular to astrophysical scales’ when work
on this paper was undertaken (EPSRC grant no EP/R014604/1).}
This research was supported in part by the International Centre for Theoretical Sciences (ICTS) for the online program - Turbulence: Problems at the Interface of Mathematics and Physics (code: ICTS/TPIMP2020/12).
NBP and RP thank the Science and Engineering Research Board (SERB) and the National Supercomputing Mission (NSM), India for support, and the Supercomputer Education and Research Centre (IISc) for computational resources.
We thank NEC, India for trial use of the SX-Aurora TSUBASA computer, on which we carried out preliminary DNSs for the 2D CHNS system. DV acknowledges the support of the Indo–French Centre for the Promotion of Advanced Scientific Research (IFCPAR/CEFIPRA) (project no. 6704-1).
\end{acknowledgements}
\providecommand{\noopsort}[1]{}\providecommand{\singleletter}[1]{#1}%
\begin{figure}[p]
  \includegraphics[width=\textwidth]{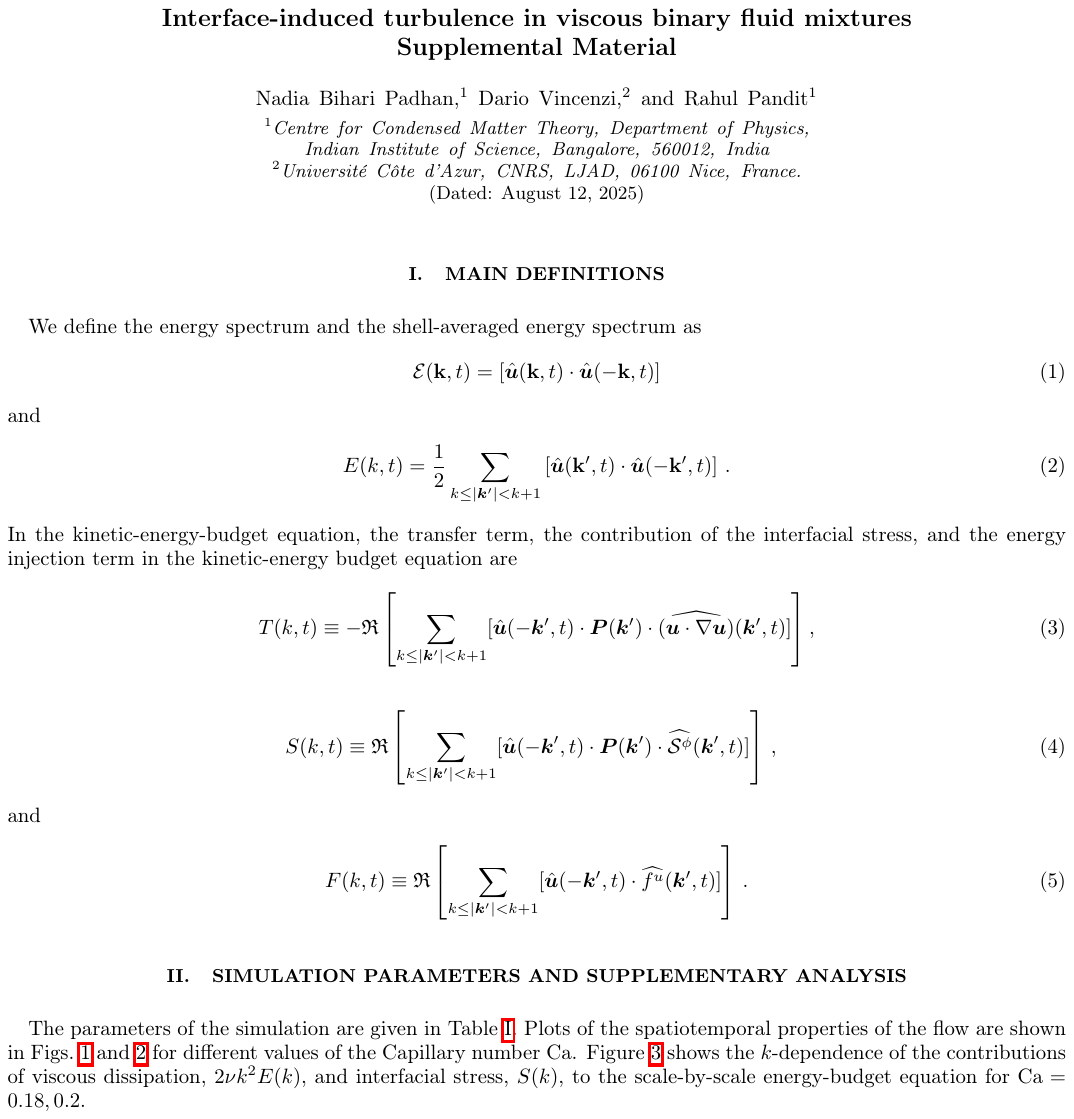}
\end{figure}
\begin{figure}[p]
  \includegraphics[width=\textwidth]{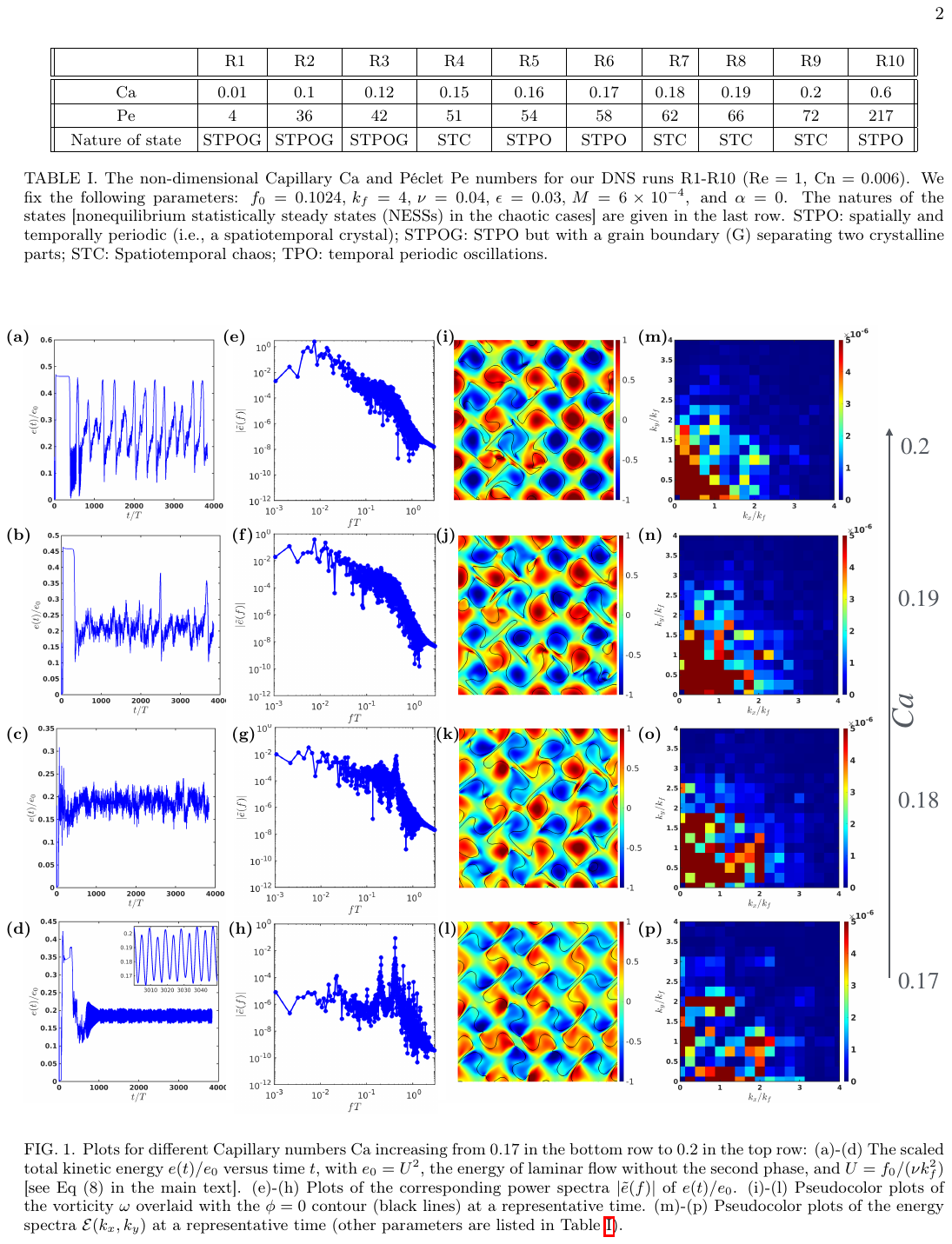}
\end{figure}
\begin{figure}[p]
  \includegraphics[width=\textwidth]{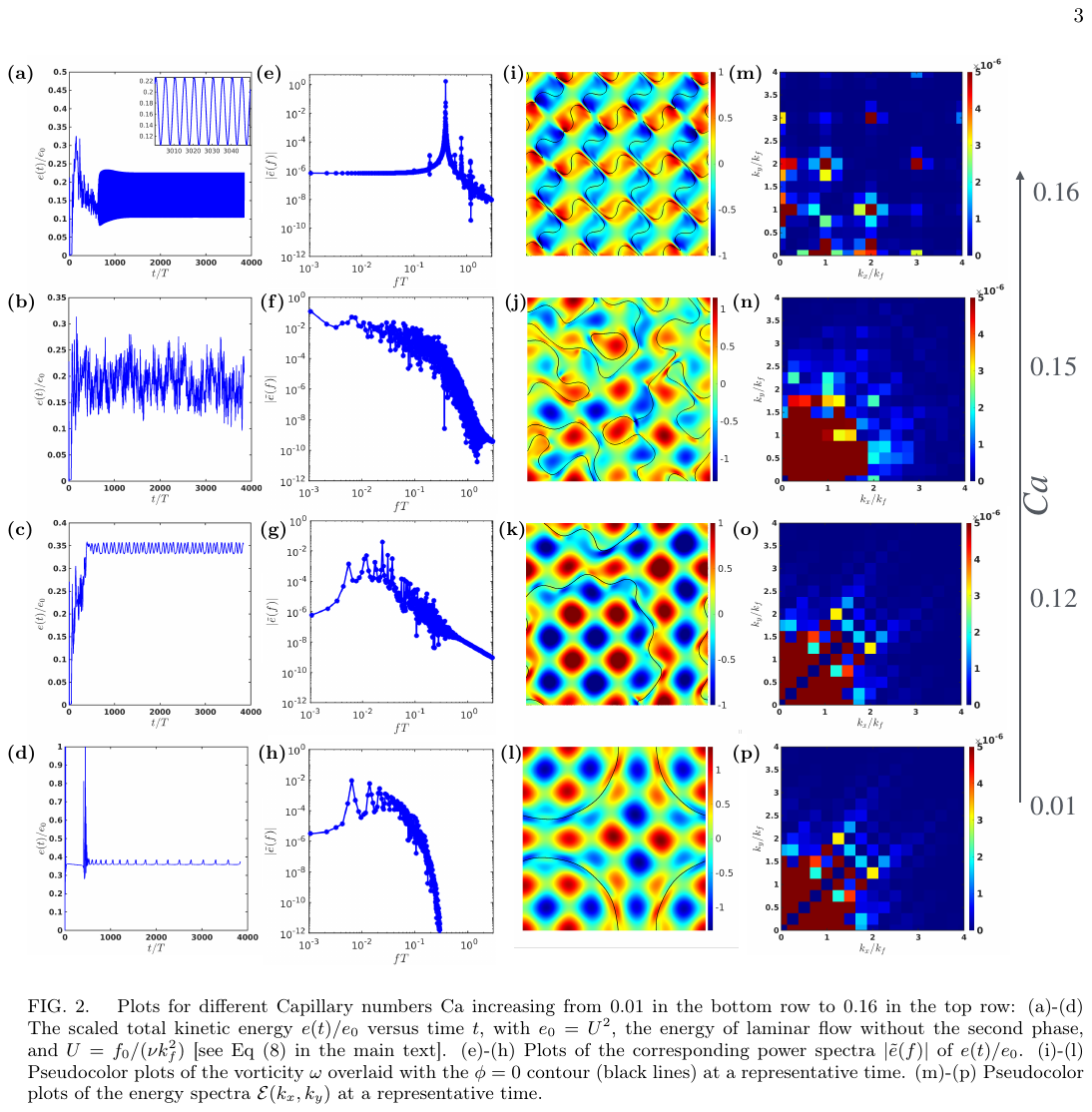}
\end{figure}
\begin{figure}[p]
  \includegraphics[width=\textwidth]{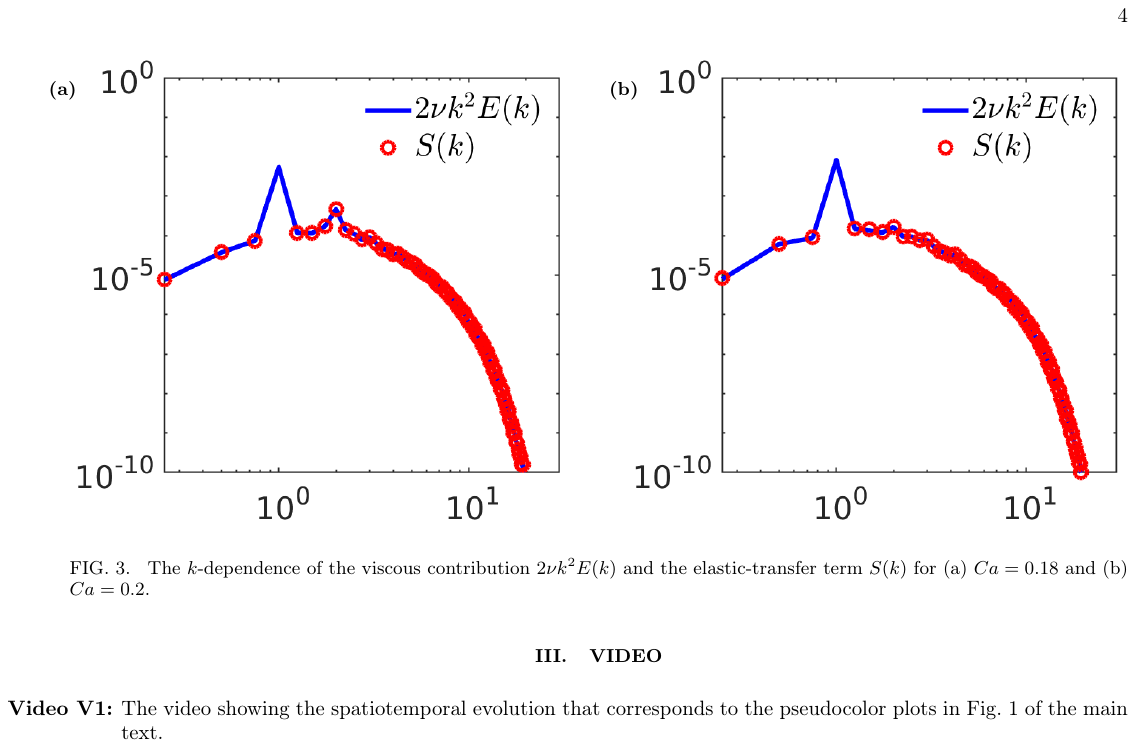}
\end{figure}

\begin{thebibliography}{51}%
\makeatletter
\providecommand \@ifxundefined [1]{%
 \@ifx{#1\undefined}
}%
\providecommand \@ifnum [1]{%
 \ifnum #1\expandafter \@firstoftwo
 \else \expandafter \@secondoftwo
 \fi
}%
\providecommand \@ifx [1]{%
 \ifx #1\expandafter \@firstoftwo
 \else \expandafter \@secondoftwo
 \fi
}%
\providecommand \natexlab [1]{#1}%
\providecommand \enquote  [1]{``#1''}%
\providecommand \bibnamefont  [1]{#1}%
\providecommand \bibfnamefont [1]{#1}%
\providecommand \citenamefont [1]{#1}%
\providecommand \href@noop [0]{\@secondoftwo}%
\providecommand \href [0]{\begingroup \@sanitize@url \@href}%
\providecommand \@href[1]{\@@startlink{#1}\@@href}%
\providecommand \@@href[1]{\endgroup#1\@@endlink}%
\providecommand \@sanitize@url [0]{\catcode `\\12\catcode `\$12\catcode `\&12\catcode `\#12\catcode `\^12\catcode `\_12\catcode `\%12\relax}%
\providecommand \@@startlink[1]{}%
\providecommand \@@endlink[0]{}%
\providecommand \url  [0]{\begingroup\@sanitize@url \@url }%
\providecommand \@url [1]{\endgroup\@href {#1}{\urlprefix }}%
\providecommand \urlprefix  [0]{URL }%
\providecommand \Eprint [0]{\href }%
\providecommand \doibase [0]{https://doi.org/}%
\providecommand \selectlanguage [0]{\@gobble}%
\providecommand \bibinfo  [0]{\@secondoftwo}%
\providecommand \bibfield  [0]{\@secondoftwo}%
\providecommand \translation [1]{[#1]}%
\providecommand \BibitemOpen [0]{}%
\providecommand \bibitemStop [0]{}%
\providecommand \bibitemNoStop [0]{.\EOS\space}%
\providecommand \EOS [0]{\spacefactor3000\relax}%
\providecommand \BibitemShut  [1]{\csname bibitem#1\endcsname}%
\let\auto@bib@innerbib\@empty
\bibitem [{\citenamefont {Groisman}\ and\ \citenamefont {Steinberg}(2000)}]{groisman2000elastic}%
  \BibitemOpen
  \bibfield  {author} {\bibinfo {author} {\bibfnamefont {A.}~\bibnamefont {Groisman}}\ and\ \bibinfo {author} {\bibfnamefont {V.}~\bibnamefont {Steinberg}},\ }\bibfield  {title} {\bibinfo {title} {Elastic turbulence in a polymer solution flow},\ }\href@noop {} {\bibfield  {journal} {\bibinfo  {journal} {Nature}\ }\textbf {\bibinfo {volume} {405}},\ \bibinfo {pages} {53} (\bibinfo {year} {2000})}\BibitemShut {NoStop}%
\bibitem [{\citenamefont {Steinberg}(2021)}]{steinberg2021elastic}%
  \BibitemOpen
  \bibfield  {author} {\bibinfo {author} {\bibfnamefont {V.}~\bibnamefont {Steinberg}},\ }\bibfield  {title} {\bibinfo {title} {Elastic turbulence: an experimental view on inertialess random flow},\ }\href@noop {} {\bibfield  {journal} {\bibinfo  {journal} {Annual Review of Fluid Mechanics}\ }\textbf {\bibinfo {volume} {53}},\ \bibinfo {pages} {27} (\bibinfo {year} {2021})}\BibitemShut {NoStop}%
\bibitem [{\citenamefont {Datta}\ \emph {et~al.}(2022)\citenamefont {Datta}, \citenamefont {Ardekani}, \citenamefont {Arratia}, \citenamefont {Beris}, \citenamefont {Bischofberger}, \citenamefont {McKinley}, \citenamefont {Eggers}, \citenamefont {L{\'o}pez-Aguilar}, \citenamefont {Fielding}, \citenamefont {Frishman} \emph {et~al.}}]{datta2022perspectives}%
  \BibitemOpen
  \bibfield  {author} {\bibinfo {author} {\bibfnamefont {S.~S.}\ \bibnamefont {Datta}}, \bibinfo {author} {\bibfnamefont {A.~M.}\ \bibnamefont {Ardekani}}, \bibinfo {author} {\bibfnamefont {P.~E.}\ \bibnamefont {Arratia}}, \bibinfo {author} {\bibfnamefont {A.~N.}\ \bibnamefont {Beris}}, \bibinfo {author} {\bibfnamefont {I.}~\bibnamefont {Bischofberger}}, \bibinfo {author} {\bibfnamefont {G.~H.}\ \bibnamefont {McKinley}}, \bibinfo {author} {\bibfnamefont {J.~G.}\ \bibnamefont {Eggers}}, \bibinfo {author} {\bibfnamefont {J.~E.}\ \bibnamefont {L{\'o}pez-Aguilar}}, \bibinfo {author} {\bibfnamefont {S.~M.}\ \bibnamefont {Fielding}}, \bibinfo {author} {\bibfnamefont {A.}~\bibnamefont {Frishman}}, \emph {et~al.},\ }\bibfield  {title} {\bibinfo {title} {Perspectives on viscoelastic flow instabilities and elastic turbulence},\ }\href@noop {} {\bibfield  {journal} {\bibinfo  {journal} {Physical Review Fluids}\ }\textbf {\bibinfo {volume} {7}},\ \bibinfo {pages} {080701} (\bibinfo {year} {2022})}\BibitemShut {NoStop}%
\bibitem [{\citenamefont {Fouxon}\ and\ \citenamefont {Lebedev}(2007)}]{fouxon2007spectra}%
  \BibitemOpen
  \bibfield  {author} {\bibinfo {author} {\bibfnamefont {A.}~\bibnamefont {Fouxon}}\ and\ \bibinfo {author} {\bibfnamefont {V.}~\bibnamefont {Lebedev}},\ }\bibfield  {title} {\bibinfo {title} {Spectra of turbulence in dilute polymer solutions},\ }\href@noop {} {\bibfield  {journal} {\bibinfo  {journal} {Physics of Fluids}\ }\textbf {\bibinfo {volume} {15}},\ \bibinfo {pages} {2060} (\bibinfo {year} {2007})}\BibitemShut {NoStop}%
\bibitem [{\citenamefont {Steinberg}(2019)}]{steinberg19}%
  \BibitemOpen
  \bibfield  {author} {\bibinfo {author} {\bibfnamefont {V.}~\bibnamefont {Steinberg}},\ }\bibfield  {title} {\bibinfo {title} {Scaling relations in elastic turbulence},\ }\href@noop {} {\bibfield  {journal} {\bibinfo  {journal} {Physical Review Letters}\ }\textbf {\bibinfo {volume} {123}},\ \bibinfo {pages} {234501} (\bibinfo {year} {2019})}\BibitemShut {NoStop}%
\bibitem [{\citenamefont {Jun}\ and\ \citenamefont {Steinberg}(2009)}]{jun2009power}%
  \BibitemOpen
  \bibfield  {author} {\bibinfo {author} {\bibfnamefont {Y.}~\bibnamefont {Jun}}\ and\ \bibinfo {author} {\bibfnamefont {V.}~\bibnamefont {Steinberg}},\ }\bibfield  {title} {\bibinfo {title} {Power and pressure fluctuations in elastic turbulence over a wide range of polymer concentrations},\ }\href@noop {} {\bibfield  {journal} {\bibinfo  {journal} {Physical Review Letters}\ }\textbf {\bibinfo {volume} {102}},\ \bibinfo {pages} {124503} (\bibinfo {year} {2009})}\BibitemShut {NoStop}%
\bibitem [{\citenamefont {Singh}\ \emph {et~al.}(2024)\citenamefont {Singh}, \citenamefont {Perlekar}, \citenamefont {Mitra},\ and\ \citenamefont {Rosti}}]{singh2024intermittency}%
  \BibitemOpen
  \bibfield  {author} {\bibinfo {author} {\bibfnamefont {R.~K.}\ \bibnamefont {Singh}}, \bibinfo {author} {\bibfnamefont {P.}~\bibnamefont {Perlekar}}, \bibinfo {author} {\bibfnamefont {D.}~\bibnamefont {Mitra}},\ and\ \bibinfo {author} {\bibfnamefont {M.~E.}\ \bibnamefont {Rosti}},\ }\bibfield  {title} {\bibinfo {title} {Intermittency in the {\itshape not-so-smooth} elastic turbulence},\ }\href@noop {} {\bibfield  {journal} {\bibinfo  {journal} {Nature Communications}\ }\textbf {\bibinfo {volume} {15}},\ \bibinfo {pages} {4070} (\bibinfo {year} {2024})}\BibitemShut {NoStop}%
\bibitem [{\citenamefont {Fardin}\ \emph {et~al.}(2010)\citenamefont {Fardin}, \citenamefont {Lopez}, \citenamefont {Croso}, \citenamefont {Gr\'egoire}, \citenamefont {Cardoso}, \citenamefont {McKinley},\ and\ \citenamefont {Lerouge}}]{fardin2010elastic}%
  \BibitemOpen
  \bibfield  {author} {\bibinfo {author} {\bibfnamefont {M.~A.}\ \bibnamefont {Fardin}}, \bibinfo {author} {\bibfnamefont {D.}~\bibnamefont {Lopez}}, \bibinfo {author} {\bibfnamefont {J.}~\bibnamefont {Croso}}, \bibinfo {author} {\bibfnamefont {G.}~\bibnamefont {Gr\'egoire}}, \bibinfo {author} {\bibfnamefont {O.}~\bibnamefont {Cardoso}}, \bibinfo {author} {\bibfnamefont {G.~H.}\ \bibnamefont {McKinley}},\ and\ \bibinfo {author} {\bibfnamefont {S.}~\bibnamefont {Lerouge}},\ }\bibfield  {title} {\bibinfo {title} {Elastic turbulence in shear banding wormlike micelles},\ }\href@noop {} {\bibfield  {journal} {\bibinfo  {journal} {Physical Review Letters}\ }\textbf {\bibinfo {volume} {104}},\ \bibinfo {pages} {178303} (\bibinfo {year} {2010})}\BibitemShut {NoStop}%
\bibitem [{\citenamefont {Majumdar}\ and\ \citenamefont {Sood}(2011)}]{majumdar2011universality}%
  \BibitemOpen
  \bibfield  {author} {\bibinfo {author} {\bibfnamefont {S.}~\bibnamefont {Majumdar}}\ and\ \bibinfo {author} {\bibfnamefont {A.~K.}\ \bibnamefont {Sood}},\ }\bibfield  {title} {\bibinfo {title} {Universality and scaling behavior of injected power in elastic turbulence in wormlike micellar gel},\ }\href@noop {} {\bibfield  {journal} {\bibinfo  {journal} {Physical Review E}\ }\textbf {\bibinfo {volume} {84}},\ \bibinfo {pages} {015302(R)} (\bibinfo {year} {2011})}\BibitemShut {NoStop}%
\bibitem [{\citenamefont {Plan}\ \emph {et~al.}(2017{\natexlab{a}})\citenamefont {Plan}, \citenamefont {Musacchio},\ and\ \citenamefont {Vincenzi}}]{emmanuel2017emergence}%
  \BibitemOpen
  \bibfield  {author} {\bibinfo {author} {\bibfnamefont {E.~L. C. V.~M.}\ \bibnamefont {Plan}}, \bibinfo {author} {\bibfnamefont {S.}~\bibnamefont {Musacchio}},\ and\ \bibinfo {author} {\bibfnamefont {D.}~\bibnamefont {Vincenzi}},\ }\bibfield  {title} {\bibinfo {title} {Emergence of chaos in a viscous solution of rods},\ }\href@noop {} {\bibfield  {journal} {\bibinfo  {journal} {Physical Review E}\ }\textbf {\bibinfo {volume} {96}},\ \bibinfo {pages} {053108} (\bibinfo {year} {2017}{\natexlab{a}})}\BibitemShut {NoStop}%
\bibitem [{\citenamefont {Puggioni}\ \emph {et~al.}(2022)\citenamefont {Puggioni}, \citenamefont {Boffetta},\ and\ \citenamefont {Musacchio}}]{puggioni2022enhancement}%
  \BibitemOpen
  \bibfield  {author} {\bibinfo {author} {\bibfnamefont {L.}~\bibnamefont {Puggioni}}, \bibinfo {author} {\bibfnamefont {G.}~\bibnamefont {Boffetta}},\ and\ \bibinfo {author} {\bibfnamefont {S.}~\bibnamefont {Musacchio}},\ }\bibfield  {title} {\bibinfo {title} {Enhancement of drag and mixing in a dilute solution of rodlike polymers at low {R}eynolds numbers},\ }\href@noop {} {\bibfield  {journal} {\bibinfo  {journal} {Physical Review Fluids}\ }\textbf {\bibinfo {volume} {7}},\ \bibinfo {pages} {083301} (\bibinfo {year} {2022})}\BibitemShut {NoStop}%
\bibitem [{\citenamefont {Puggioni}\ and\ \citenamefont {Musacchio}(2024)}]{PhysRevE.110.015104}%
  \BibitemOpen
  \bibfield  {author} {\bibinfo {author} {\bibfnamefont {L.}~\bibnamefont {Puggioni}}\ and\ \bibinfo {author} {\bibfnamefont {S.}~\bibnamefont {Musacchio}},\ }\bibfield  {title} {\bibinfo {title} {Orientational order and topological defects in a dilute solutions of rodlike polymers at low reynolds number},\ }\href {https://doi.org/10.1103/PhysRevE.110.015104} {\bibfield  {journal} {\bibinfo  {journal} {Phys. Rev. E}\ }\textbf {\bibinfo {volume} {110}},\ \bibinfo {pages} {015104} (\bibinfo {year} {2024})}\BibitemShut {NoStop}%
\bibitem [{\citenamefont {Souzy}\ \emph {et~al.}(2017)\citenamefont {Souzy}, \citenamefont {Lhuissier}, \citenamefont {Villermaux},\ and\ \citenamefont {Metzger}}]{souzy2017stretching}%
  \BibitemOpen
  \bibfield  {author} {\bibinfo {author} {\bibfnamefont {M.}~\bibnamefont {Souzy}}, \bibinfo {author} {\bibfnamefont {H.}~\bibnamefont {Lhuissier}}, \bibinfo {author} {\bibfnamefont {E.}~\bibnamefont {Villermaux}},\ and\ \bibinfo {author} {\bibfnamefont {B.}~\bibnamefont {Metzger}},\ }\bibfield  {title} {\bibinfo {title} {Stretching and mixing in sheared particulate suspensions},\ }\href@noop {} {\bibfield  {journal} {\bibinfo  {journal} {Journal of Fluid Mechanics}\ }\textbf {\bibinfo {volume} {812}},\ \bibinfo {pages} {611} (\bibinfo {year} {2017})}\BibitemShut {NoStop}%
\bibitem [{\citenamefont {Turuban}\ \emph {et~al.}(2021)\citenamefont {Turuban}, \citenamefont {Lhuissier},\ and\ \citenamefont {Metzger}}]{turuban2021mixing}%
  \BibitemOpen
  \bibfield  {author} {\bibinfo {author} {\bibfnamefont {R.}~\bibnamefont {Turuban}}, \bibinfo {author} {\bibfnamefont {H.}~\bibnamefont {Lhuissier}},\ and\ \bibinfo {author} {\bibfnamefont {B.}~\bibnamefont {Metzger}},\ }\bibfield  {title} {\bibinfo {title} {Mixing in a sheared particulate suspension},\ }\href@noop {} {\bibfield  {journal} {\bibinfo  {journal} {Journal of Fluid Mechanics}\ }\textbf {\bibinfo {volume} {916}},\ \bibinfo {pages} {R4} (\bibinfo {year} {2021})}\BibitemShut {NoStop}%
\bibitem [{\citenamefont {Frisch}(1995)}]{frisch1995turbulence}%
  \BibitemOpen
  \bibfield  {author} {\bibinfo {author} {\bibfnamefont {U.}~\bibnamefont {Frisch}},\ }\href@noop {} {\emph {\bibinfo {title} {Turbulence: The Legacy of A.~N.~Kolmogorov}}}\ (\bibinfo  {publisher} {Cambridge University Press, Cambridge, UK},\ \bibinfo {year} {1995})\BibitemShut {NoStop}%
\bibitem [{\citenamefont {Groisman}\ and\ \citenamefont {Steinberg}(2001)}]{groisman01}%
  \BibitemOpen
  \bibfield  {author} {\bibinfo {author} {\bibfnamefont {A.}~\bibnamefont {Groisman}}\ and\ \bibinfo {author} {\bibfnamefont {V.}~\bibnamefont {Steinberg}},\ }\bibfield  {title} {\bibinfo {title} {Efficient mixing at low {R}eynolds numbers using polymer additives},\ }\href@noop {} {\bibfield  {journal} {\bibinfo  {journal} {Nature}\ }\textbf {\bibinfo {volume} {405}},\ \bibinfo {pages} {905–908} (\bibinfo {year} {2001})}\BibitemShut {NoStop}%
\bibitem [{\citenamefont {Aref}\ \emph {et~al.}(2017)\citenamefont {Aref}, \citenamefont {Blake}, \citenamefont {Budi{\v{s}}i{\'c}}, \citenamefont {Cardoso}, \citenamefont {Cartwright}, \citenamefont {Clercx}, \citenamefont {El~Omari}, \citenamefont {Feudel}, \citenamefont {Golestanian}, \citenamefont {Gouillart} \emph {et~al.}}]{aref2017frontiers}%
  \BibitemOpen
  \bibfield  {author} {\bibinfo {author} {\bibfnamefont {H.}~\bibnamefont {Aref}}, \bibinfo {author} {\bibfnamefont {J.~R.}\ \bibnamefont {Blake}}, \bibinfo {author} {\bibfnamefont {M.}~\bibnamefont {Budi{\v{s}}i{\'c}}}, \bibinfo {author} {\bibfnamefont {S.~S.}\ \bibnamefont {Cardoso}}, \bibinfo {author} {\bibfnamefont {J.~H.}\ \bibnamefont {Cartwright}}, \bibinfo {author} {\bibfnamefont {H.~J.}\ \bibnamefont {Clercx}}, \bibinfo {author} {\bibfnamefont {K.}~\bibnamefont {El~Omari}}, \bibinfo {author} {\bibfnamefont {U.}~\bibnamefont {Feudel}}, \bibinfo {author} {\bibfnamefont {R.}~\bibnamefont {Golestanian}}, \bibinfo {author} {\bibfnamefont {E.}~\bibnamefont {Gouillart}}, \emph {et~al.},\ }\bibfield  {title} {\bibinfo {title} {Frontiers of chaotic advection},\ }\href@noop {} {\bibfield  {journal} {\bibinfo  {journal} {Reviews of Modern Physics}\ }\textbf {\bibinfo {volume} {89}},\ \bibinfo {pages} {025007} (\bibinfo {year} {2017})}\BibitemShut {NoStop}%
\bibitem [{\citenamefont {Ho}\ \emph {et~al.}(2022)\citenamefont {Ho}, \citenamefont {Razzaghi}, \citenamefont {Ramachandran},\ and\ \citenamefont {Mikkonen}}]{ho2022emulsion}%
  \BibitemOpen
  \bibfield  {author} {\bibinfo {author} {\bibfnamefont {T.~M.}\ \bibnamefont {Ho}}, \bibinfo {author} {\bibfnamefont {A.}~\bibnamefont {Razzaghi}}, \bibinfo {author} {\bibfnamefont {A.}~\bibnamefont {Ramachandran}},\ and\ \bibinfo {author} {\bibfnamefont {K.~S.}\ \bibnamefont {Mikkonen}},\ }\bibfield  {title} {\bibinfo {title} {Emulsion characterization via microfluidic devices: A review on interfacial tension and stability to coalescence},\ }\href@noop {} {\bibfield  {journal} {\bibinfo  {journal} {Advances in Colloid and Interface Science}\ }\textbf {\bibinfo {volume} {299}},\ \bibinfo {pages} {102541} (\bibinfo {year} {2022})}\BibitemShut {NoStop}%
\bibitem [{\citenamefont {Gunes}(2018)}]{gunes2018microfluidics}%
  \BibitemOpen
  \bibfield  {author} {\bibinfo {author} {\bibfnamefont {D.~Z.}\ \bibnamefont {Gunes}},\ }\bibfield  {title} {\bibinfo {title} {Microfluidics for food science and engineering},\ }\href@noop {} {\bibfield  {journal} {\bibinfo  {journal} {Current Opinion in Food Science}\ }\textbf {\bibinfo {volume} {21}},\ \bibinfo {pages} {57} (\bibinfo {year} {2018})}\BibitemShut {NoStop}%
\bibitem [{\citenamefont {Gilbert}\ \emph {et~al.}(2013)\citenamefont {Gilbert}, \citenamefont {Picard}, \citenamefont {Savary},\ and\ \citenamefont {Grisel}}]{gilbert2013rheological}%
  \BibitemOpen
  \bibfield  {author} {\bibinfo {author} {\bibfnamefont {L.}~\bibnamefont {Gilbert}}, \bibinfo {author} {\bibfnamefont {C.}~\bibnamefont {Picard}}, \bibinfo {author} {\bibfnamefont {G.}~\bibnamefont {Savary}},\ and\ \bibinfo {author} {\bibfnamefont {M.}~\bibnamefont {Grisel}},\ }\bibfield  {title} {\bibinfo {title} {Rheological and textural characterization of cosmetic emulsions containing natural and synthetic polymers: relationships between both data},\ }\href@noop {} {\bibfield  {journal} {\bibinfo  {journal} {Colloids and Surfaces A: Physicochemical and Engineering Aspects}\ }\textbf {\bibinfo {volume} {421}},\ \bibinfo {pages} {150} (\bibinfo {year} {2013})}\BibitemShut {NoStop}%
\bibitem [{\citenamefont {Maeki}(2019)}]{maeki2019microfluidics}%
  \BibitemOpen
  \bibfield  {author} {\bibinfo {author} {\bibfnamefont {M.}~\bibnamefont {Maeki}},\ }\bibfield  {title} {\bibinfo {title} {Microfluidics for pharmaceutical applications},\ }in\ \href@noop {} {\emph {\bibinfo {booktitle} {Microfluidics for Pharmaceutical Applications}}}\ (\bibinfo  {publisher} {Elsevier},\ \bibinfo {year} {2019})\ pp.\ \bibinfo {pages} {101--119}\BibitemShut {NoStop}%
\bibitem [{\citenamefont {Zhao}(2013)}]{zhao2013multiphase}%
  \BibitemOpen
  \bibfield  {author} {\bibinfo {author} {\bibfnamefont {C.-X.}\ \bibnamefont {Zhao}},\ }\bibfield  {title} {\bibinfo {title} {Multiphase flow microfluidics for the production of single or multiple emulsions for drug delivery},\ }\href@noop {} {\bibfield  {journal} {\bibinfo  {journal} {Advanced Drug Delivery Reviews}\ }\textbf {\bibinfo {volume} {65}},\ \bibinfo {pages} {1420} (\bibinfo {year} {2013})}\BibitemShut {NoStop}%
\bibitem [{\citenamefont {Scarbolo}\ \emph {et~al.}(2015)\citenamefont {Scarbolo}, \citenamefont {Bianco},\ and\ \citenamefont {Soldati}}]{scarbolo2015coalescence}%
  \BibitemOpen
  \bibfield  {author} {\bibinfo {author} {\bibfnamefont {L.}~\bibnamefont {Scarbolo}}, \bibinfo {author} {\bibfnamefont {F.}~\bibnamefont {Bianco}},\ and\ \bibinfo {author} {\bibfnamefont {A.}~\bibnamefont {Soldati}},\ }\bibfield  {title} {\bibinfo {title} {Coalescence and breakup of large droplets in turbulent channel flow},\ }\href@noop {} {\bibfield  {journal} {\bibinfo  {journal} {Physics of Fluids}\ }\textbf {\bibinfo {volume} {27}},\ \bibinfo {pages} {073302} (\bibinfo {year} {2015})}\BibitemShut {NoStop}%
\bibitem [{\citenamefont {Pal}\ \emph {et~al.}(2016)\citenamefont {Pal}, \citenamefont {Perlekar}, \citenamefont {Gupta},\ and\ \citenamefont {Pandit}}]{pal2016binary}%
  \BibitemOpen
  \bibfield  {author} {\bibinfo {author} {\bibfnamefont {N.}~\bibnamefont {Pal}}, \bibinfo {author} {\bibfnamefont {P.}~\bibnamefont {Perlekar}}, \bibinfo {author} {\bibfnamefont {A.}~\bibnamefont {Gupta}},\ and\ \bibinfo {author} {\bibfnamefont {R.}~\bibnamefont {Pandit}},\ }\bibfield  {title} {\bibinfo {title} {Binary-fluid turbulence: Signatures of multifractal droplet dynamics and dissipation reduction},\ }\href@noop {} {\bibfield  {journal} {\bibinfo  {journal} {Physical Review E}\ }\textbf {\bibinfo {volume} {93}},\ \bibinfo {pages} {063115} (\bibinfo {year} {2016})}\BibitemShut {NoStop}%
\bibitem [{\citenamefont {Roccon}\ \emph {et~al.}(2017)\citenamefont {Roccon}, \citenamefont {De~Paoli}, \citenamefont {Zonta},\ and\ \citenamefont {Soldati}}]{roccon2017viscosity}%
  \BibitemOpen
  \bibfield  {author} {\bibinfo {author} {\bibfnamefont {A.}~\bibnamefont {Roccon}}, \bibinfo {author} {\bibfnamefont {M.}~\bibnamefont {De~Paoli}}, \bibinfo {author} {\bibfnamefont {F.}~\bibnamefont {Zonta}},\ and\ \bibinfo {author} {\bibfnamefont {A.}~\bibnamefont {Soldati}},\ }\bibfield  {title} {\bibinfo {title} {Viscosity-modulated breakup and coalescence of large drops in bounded turbulence},\ }\href@noop {} {\bibfield  {journal} {\bibinfo  {journal} {Physical Review Fluids}\ }\textbf {\bibinfo {volume} {2}},\ \bibinfo {pages} {083603} (\bibinfo {year} {2017})}\BibitemShut {NoStop}%
\bibitem [{\citenamefont {Negro}\ \emph {et~al.}(2023)\citenamefont {Negro}, \citenamefont {Carenza}, \citenamefont {Gonnella}, \citenamefont {Mackay}, \citenamefont {Morozov},\ and\ \citenamefont {Marenduzzo}}]{negro2023yield}%
  \BibitemOpen
  \bibfield  {author} {\bibinfo {author} {\bibfnamefont {G.}~\bibnamefont {Negro}}, \bibinfo {author} {\bibfnamefont {L.~N.}\ \bibnamefont {Carenza}}, \bibinfo {author} {\bibfnamefont {G.}~\bibnamefont {Gonnella}}, \bibinfo {author} {\bibfnamefont {F.}~\bibnamefont {Mackay}}, \bibinfo {author} {\bibfnamefont {A.}~\bibnamefont {Morozov}},\ and\ \bibinfo {author} {\bibfnamefont {D.}~\bibnamefont {Marenduzzo}},\ }\bibfield  {title} {\bibinfo {title} {Yield-stress transition in suspensions of deformable droplets},\ }\href@noop {} {\bibfield  {journal} {\bibinfo  {journal} {Science Advances}\ }\textbf {\bibinfo {volume} {9}},\ \bibinfo {pages} {eadf8106} (\bibinfo {year} {2023})}\BibitemShut {NoStop}%
\bibitem [{\citenamefont {Elghobashi}(2019)}]{elghobashi2019direct}%
  \BibitemOpen
  \bibfield  {author} {\bibinfo {author} {\bibfnamefont {S.}~\bibnamefont {Elghobashi}},\ }\bibfield  {title} {\bibinfo {title} {Direct numerical simulation of turbulent flows laden with droplets or bubbles},\ }\href@noop {} {\bibfield  {journal} {\bibinfo  {journal} {Annual Review of Fluid Mechanics}\ }\textbf {\bibinfo {volume} {51}},\ \bibinfo {pages} {217} (\bibinfo {year} {2019})}\BibitemShut {NoStop}%
\bibitem [{\citenamefont {Pal}\ \emph {et~al.}(2022)\citenamefont {Pal}, \citenamefont {Ramadugu}, \citenamefont {Perlekar},\ and\ \citenamefont {Pandit}}]{pal2022ephemeral}%
  \BibitemOpen
  \bibfield  {author} {\bibinfo {author} {\bibfnamefont {N.}~\bibnamefont {Pal}}, \bibinfo {author} {\bibfnamefont {R.}~\bibnamefont {Ramadugu}}, \bibinfo {author} {\bibfnamefont {P.}~\bibnamefont {Perlekar}},\ and\ \bibinfo {author} {\bibfnamefont {R.}~\bibnamefont {Pandit}},\ }\bibfield  {title} {\bibinfo {title} {Ephemeral antibubbles: Spatiotemporal evolution from direct numerical simulations},\ }\href@noop {} {\bibfield  {journal} {\bibinfo  {journal} {Physical Review Research}\ }\textbf {\bibinfo {volume} {4}},\ \bibinfo {pages} {043128} (\bibinfo {year} {2022})}\BibitemShut {NoStop}%
\bibitem [{\citenamefont {Perlekar}\ \emph {et~al.}(2017)\citenamefont {Perlekar}, \citenamefont {Pal},\ and\ \citenamefont {Pandit}}]{perlekar2017two}%
  \BibitemOpen
  \bibfield  {author} {\bibinfo {author} {\bibfnamefont {P.}~\bibnamefont {Perlekar}}, \bibinfo {author} {\bibfnamefont {N.}~\bibnamefont {Pal}},\ and\ \bibinfo {author} {\bibfnamefont {R.}~\bibnamefont {Pandit}},\ }\bibfield  {title} {\bibinfo {title} {Two-dimensional turbulence in symmetric binary-fluid mixtures: Coarsening arrest by the inverse cascade},\ }\href@noop {} {\bibfield  {journal} {\bibinfo  {journal} {Scientific Reports}\ }\textbf {\bibinfo {volume} {7}},\ \bibinfo {pages} {44589} (\bibinfo {year} {2017})}\BibitemShut {NoStop}%
\bibitem [{\citenamefont {Shek}\ and\ \citenamefont {Kusumaatmaja}(2022)}]{shek2022spontaneous}%
  \BibitemOpen
  \bibfield  {author} {\bibinfo {author} {\bibfnamefont {A.~C.}\ \bibnamefont {Shek}}\ and\ \bibinfo {author} {\bibfnamefont {H.}~\bibnamefont {Kusumaatmaja}},\ }\bibfield  {title} {\bibinfo {title} {Spontaneous phase separation of ternary fluid mixtures},\ }\href@noop {} {\bibfield  {journal} {\bibinfo  {journal} {Soft Matter}\ }\textbf {\bibinfo {volume} {18}},\ \bibinfo {pages} {5807} (\bibinfo {year} {2022})}\BibitemShut {NoStop}%
\bibitem [{\citenamefont {Perlekar}\ \emph {et~al.}(2014)\citenamefont {Perlekar}, \citenamefont {Benzi}, \citenamefont {Clercx}, \citenamefont {Nelson},\ and\ \citenamefont {Toschi}}]{perlekar2014spinodal}%
  \BibitemOpen
  \bibfield  {author} {\bibinfo {author} {\bibfnamefont {P.}~\bibnamefont {Perlekar}}, \bibinfo {author} {\bibfnamefont {R.}~\bibnamefont {Benzi}}, \bibinfo {author} {\bibfnamefont {H.~J.}\ \bibnamefont {Clercx}}, \bibinfo {author} {\bibfnamefont {D.~R.}\ \bibnamefont {Nelson}},\ and\ \bibinfo {author} {\bibfnamefont {F.}~\bibnamefont {Toschi}},\ }\bibfield  {title} {\bibinfo {title} {Spinodal decomposition in homogeneous and isotropic turbulence},\ }\href@noop {} {\bibfield  {journal} {\bibinfo  {journal} {Physical Review Letters}\ }\textbf {\bibinfo {volume} {112}},\ \bibinfo {pages} {014502} (\bibinfo {year} {2014})}\BibitemShut {NoStop}%
\bibitem [{\citenamefont {Fan}\ \emph {et~al.}(2016)\citenamefont {Fan}, \citenamefont {Diamond}, \citenamefont {Chac{\'o}n},\ and\ \citenamefont {Li}}]{fan2016cascades}%
  \BibitemOpen
  \bibfield  {author} {\bibinfo {author} {\bibfnamefont {X.}~\bibnamefont {Fan}}, \bibinfo {author} {\bibfnamefont {P.}~\bibnamefont {Diamond}}, \bibinfo {author} {\bibfnamefont {L.}~\bibnamefont {Chac{\'o}n}},\ and\ \bibinfo {author} {\bibfnamefont {H.}~\bibnamefont {Li}},\ }\bibfield  {title} {\bibinfo {title} {Cascades and spectra of a turbulent spinodal decomposition in two-dimensional symmetric binary liquid mixtures},\ }\href@noop {} {\bibfield  {journal} {\bibinfo  {journal} {Physical Review Fluids}\ }\textbf {\bibinfo {volume} {1}},\ \bibinfo {pages} {054403} (\bibinfo {year} {2016})}\BibitemShut {NoStop}%
\bibitem [{\citenamefont {Fan}\ \emph {et~al.}(2018)\citenamefont {Fan}, \citenamefont {Diamond},\ and\ \citenamefont {Chac{\'o}n}}]{fan2018chns}%
  \BibitemOpen
  \bibfield  {author} {\bibinfo {author} {\bibfnamefont {X.}~\bibnamefont {Fan}}, \bibinfo {author} {\bibfnamefont {P.}~\bibnamefont {Diamond}},\ and\ \bibinfo {author} {\bibfnamefont {L.}~\bibnamefont {Chac{\'o}n}},\ }\bibfield  {title} {\bibinfo {title} {{CHNS}: A case study of turbulence in elastic media},\ }\href@noop {} {\bibfield  {journal} {\bibinfo  {journal} {Physics of Plasmas}\ }\textbf {\bibinfo {volume} {25}} (\bibinfo {year} {2018})}\BibitemShut {NoStop}%
\bibitem [{\citenamefont {Perlekar}\ and\ \citenamefont {Pandit}(2010)}]{perlekar2010turbulence}%
  \BibitemOpen
  \bibfield  {author} {\bibinfo {author} {\bibfnamefont {P.}~\bibnamefont {Perlekar}}\ and\ \bibinfo {author} {\bibfnamefont {R.}~\bibnamefont {Pandit}},\ }\bibfield  {title} {\bibinfo {title} {Turbulence-induced melting of a nonequilibrium vortex crystal in a forced thin fluid film},\ }\href@noop {} {\bibfield  {journal} {\bibinfo  {journal} {New Journal of Physics}\ }\textbf {\bibinfo {volume} {12}},\ \bibinfo {pages} {023033} (\bibinfo {year} {2010})}\BibitemShut {NoStop}%
\bibitem [{\citenamefont {Gupta}\ and\ \citenamefont {Pandit}(2017)}]{gupta2017melting}%
  \BibitemOpen
  \bibfield  {author} {\bibinfo {author} {\bibfnamefont {A.}~\bibnamefont {Gupta}}\ and\ \bibinfo {author} {\bibfnamefont {R.}~\bibnamefont {Pandit}},\ }\bibfield  {title} {\bibinfo {title} {Melting of a nonequilibrium vortex crystal in a fluid film with polymers: Elastic versus fluid turbulence},\ }\href@noop {} {\bibfield  {journal} {\bibinfo  {journal} {Physical Review E}\ }\textbf {\bibinfo {volume} {95}},\ \bibinfo {pages} {033119} (\bibinfo {year} {2017})}\BibitemShut {NoStop}%
\bibitem [{\citenamefont {Plan}\ \emph {et~al.}(2017{\natexlab{b}})\citenamefont {Plan}, \citenamefont {Gupta}, \citenamefont {Vincenzi},\ and\ \citenamefont {Gibbon}}]{plan2017lyapunov}%
  \BibitemOpen
  \bibfield  {author} {\bibinfo {author} {\bibfnamefont {E.~L. C. V.~M.}\ \bibnamefont {Plan}}, \bibinfo {author} {\bibfnamefont {A.}~\bibnamefont {Gupta}}, \bibinfo {author} {\bibfnamefont {D.}~\bibnamefont {Vincenzi}},\ and\ \bibinfo {author} {\bibfnamefont {J.~D.}\ \bibnamefont {Gibbon}},\ }\bibfield  {title} {\bibinfo {title} {Lyapunov dimension of elastic turbulence},\ }\href@noop {} {\bibfield  {journal} {\bibinfo  {journal} {Journal of Fluid Mechanics}\ }\textbf {\bibinfo {volume} {822}} (\bibinfo {year} {2017}{\natexlab{b}})}\BibitemShut {NoStop}%
\bibitem [{sup()}]{supmat}%
  \BibitemOpen
  \href@noop {} {}\bibinfo {note} {See supplemental material at}\BibitemShut {NoStop}%
\bibitem [{\citenamefont {Canuto}\ \emph {et~al.}(2012)\citenamefont {Canuto}, \citenamefont {Hussaini}, \citenamefont {Quarteroni}, \citenamefont {Thomas~Jr} \emph {et~al.}}]{canuto2012spectral}%
  \BibitemOpen
  \bibfield  {author} {\bibinfo {author} {\bibfnamefont {C.}~\bibnamefont {Canuto}}, \bibinfo {author} {\bibfnamefont {M.~Y.}\ \bibnamefont {Hussaini}}, \bibinfo {author} {\bibfnamefont {A.}~\bibnamefont {Quarteroni}}, \bibinfo {author} {\bibfnamefont {A.}~\bibnamefont {Thomas~Jr}}, \emph {et~al.},\ }\href@noop {} {\emph {\bibinfo {title} {Spectral methods in fluid dynamics}}}\ (\bibinfo  {publisher} {Springer Science Business Media},\ \bibinfo {year} {2012})\BibitemShut {NoStop}%
\bibitem [{\citenamefont {Padhan}\ and\ \citenamefont {Pandit}(2023{\natexlab{a}})}]{padhan2023activity}%
  \BibitemOpen
  \bibfield  {author} {\bibinfo {author} {\bibfnamefont {N.~B.}\ \bibnamefont {Padhan}}\ and\ \bibinfo {author} {\bibfnamefont {R.}~\bibnamefont {Pandit}},\ }\bibfield  {title} {\bibinfo {title} {Activity-induced droplet propulsion and multifractality},\ }\href@noop {} {\bibfield  {journal} {\bibinfo  {journal} {Physical Review Research}\ }\textbf {\bibinfo {volume} {5}},\ \bibinfo {pages} {L032013} (\bibinfo {year} {2023}{\natexlab{a}})}\BibitemShut {NoStop}%
\bibitem [{\citenamefont {Padhan}\ and\ \citenamefont {Pandit}(2023{\natexlab{b}})}]{padhan2023unveiling}%
  \BibitemOpen
  \bibfield  {author} {\bibinfo {author} {\bibfnamefont {N.~B.}\ \bibnamefont {Padhan}}\ and\ \bibinfo {author} {\bibfnamefont {R.}~\bibnamefont {Pandit}},\ }\bibfield  {title} {\bibinfo {title} {Unveiling the spatiotemporal evolution of liquid-lens coalescence: Self-similarity, vortex quadrupoles, and turbulence in a three-phase fluid system},\ }\href@noop {} {\bibfield  {journal} {\bibinfo  {journal} {Physics of Fluids}\ }\textbf {\bibinfo {volume} {35}} (\bibinfo {year} {2023}{\natexlab{b}})}\BibitemShut {NoStop}%
\bibitem [{\citenamefont {Cox}\ and\ \citenamefont {Matthews}(2002)}]{cox2002exponential}%
  \BibitemOpen
  \bibfield  {author} {\bibinfo {author} {\bibfnamefont {S.~M.}\ \bibnamefont {Cox}}\ and\ \bibinfo {author} {\bibfnamefont {P.~C.}\ \bibnamefont {Matthews}},\ }\bibfield  {title} {\bibinfo {title} {Exponential time differencing for stiff systems},\ }\href@noop {} {\bibfield  {journal} {\bibinfo  {journal} {Journal of Computational Physics}\ }\textbf {\bibinfo {volume} {176}},\ \bibinfo {pages} {430} (\bibinfo {year} {2002})}\BibitemShut {NoStop}%
\bibitem [{\citenamefont {Benzi}\ \emph {et~al.}(2010)\citenamefont {Benzi}, \citenamefont {Biferale}, \citenamefont {Fisher}, \citenamefont {Lamb},\ and\ \citenamefont {Toschi}}]{benzi2010inertial}%
  \BibitemOpen
  \bibfield  {author} {\bibinfo {author} {\bibfnamefont {R.}~\bibnamefont {Benzi}}, \bibinfo {author} {\bibfnamefont {L.}~\bibnamefont {Biferale}}, \bibinfo {author} {\bibfnamefont {R.}~\bibnamefont {Fisher}}, \bibinfo {author} {\bibfnamefont {D.}~\bibnamefont {Lamb}},\ and\ \bibinfo {author} {\bibfnamefont {F.}~\bibnamefont {Toschi}},\ }\bibfield  {title} {\bibinfo {title} {Inertial range {Eulerian} and {Lagrangian} statistics from numerical simulations of isotropic turbulence},\ }\href@noop {} {\bibfield  {journal} {\bibinfo  {journal} {Journal of Fluid Mechanics}\ }\textbf {\bibinfo {volume} {653}},\ \bibinfo {pages} {221} (\bibinfo {year} {2010})}\BibitemShut {NoStop}%
\bibitem [{\citenamefont {Verma}\ \emph {et~al.}(2020)\citenamefont {Verma}, \citenamefont {Bhatnagar}, \citenamefont {Mitra},\ and\ \citenamefont {Pandit}}]{verma2020first}%
  \BibitemOpen
  \bibfield  {author} {\bibinfo {author} {\bibfnamefont {A.~K.}\ \bibnamefont {Verma}}, \bibinfo {author} {\bibfnamefont {A.}~\bibnamefont {Bhatnagar}}, \bibinfo {author} {\bibfnamefont {D.}~\bibnamefont {Mitra}},\ and\ \bibinfo {author} {\bibfnamefont {R.}~\bibnamefont {Pandit}},\ }\bibfield  {title} {\bibinfo {title} {First-passage-time problem for tracers in turbulent flows applied to virus spreading},\ }\href@noop {} {\bibfield  {journal} {\bibinfo  {journal} {Physical Review Research}\ }\textbf {\bibinfo {volume} {2}},\ \bibinfo {pages} {033239} (\bibinfo {year} {2020})}\BibitemShut {NoStop}%
\bibitem [{Note1()}]{Note1}%
  \BibitemOpen
  \bibinfo {note} {We have checked explicitly that our results are independent of the initial arrangements and sizes of the droplets.}\BibitemShut {Stop}%
\bibitem [{\citenamefont {Gotoh}\ and\ \citenamefont {Yamada}(1984)}]{gotoh1984instability}%
  \BibitemOpen
  \bibfield  {author} {\bibinfo {author} {\bibfnamefont {K.}~\bibnamefont {Gotoh}}\ and\ \bibinfo {author} {\bibfnamefont {M.}~\bibnamefont {Yamada}},\ }\bibfield  {title} {\bibinfo {title} {Instability of a cellular flow},\ }\href@noop {} {\bibfield  {journal} {\bibinfo  {journal} {Journal of the Physical Society of Japan}\ }\textbf {\bibinfo {volume} {53}},\ \bibinfo {pages} {3395} (\bibinfo {year} {1984})}\BibitemShut {NoStop}%
\bibitem [{\citenamefont {Boffetta}\ and\ \citenamefont {Ecke}(2012)}]{boffetta2012two}%
  \BibitemOpen
  \bibfield  {author} {\bibinfo {author} {\bibfnamefont {G.}~\bibnamefont {Boffetta}}\ and\ \bibinfo {author} {\bibfnamefont {R.~E.}\ \bibnamefont {Ecke}},\ }\bibfield  {title} {\bibinfo {title} {Two-dimensional turbulence},\ }\href@noop {} {\bibfield  {journal} {\bibinfo  {journal} {Annual Review of Fluid Mechanics}\ }\textbf {\bibinfo {volume} {44}},\ \bibinfo {pages} {427} (\bibinfo {year} {2012})}\BibitemShut {NoStop}%
\bibitem [{\citenamefont {Pandit}\ \emph {et~al.}(2017)\citenamefont {Pandit}, \citenamefont {Banerjee}, \citenamefont {Bhatnagar}, \citenamefont {Brachet}, \citenamefont {Gupta}, \citenamefont {Mitra}, \citenamefont {Pal}, \citenamefont {Perlekar}, \citenamefont {Ray}, \citenamefont {Shukla} \emph {et~al.}}]{pandit2017overview}%
  \BibitemOpen
  \bibfield  {author} {\bibinfo {author} {\bibfnamefont {R.}~\bibnamefont {Pandit}}, \bibinfo {author} {\bibfnamefont {D.}~\bibnamefont {Banerjee}}, \bibinfo {author} {\bibfnamefont {A.}~\bibnamefont {Bhatnagar}}, \bibinfo {author} {\bibfnamefont {M.}~\bibnamefont {Brachet}}, \bibinfo {author} {\bibfnamefont {A.}~\bibnamefont {Gupta}}, \bibinfo {author} {\bibfnamefont {D.}~\bibnamefont {Mitra}}, \bibinfo {author} {\bibfnamefont {N.}~\bibnamefont {Pal}}, \bibinfo {author} {\bibfnamefont {P.}~\bibnamefont {Perlekar}}, \bibinfo {author} {\bibfnamefont {S.~S.}\ \bibnamefont {Ray}}, \bibinfo {author} {\bibfnamefont {V.}~\bibnamefont {Shukla}}, \emph {et~al.},\ }\bibfield  {title} {\bibinfo {title} {An overview of the statistical properties of two-dimensional turbulence in fluids with particles, conducting fluids, fluids with polymer additives, binary-fluid mixtures, and superfluids},\ }\href@noop {} {\bibfield  {journal} {\bibinfo  {journal} {Physics of Fluids}\ }\textbf {\bibinfo {volume} {29}} (\bibinfo {year}
  {2017})}\BibitemShut {NoStop}%
\bibitem [{\citenamefont {Perlekar}(2019)}]{perlekar2019kinetic}%
  \BibitemOpen
  \bibfield  {author} {\bibinfo {author} {\bibfnamefont {P.}~\bibnamefont {Perlekar}},\ }\bibfield  {title} {\bibinfo {title} {Kinetic energy spectra and flux in turbulent phase-separating symmetric binary-fluid mixtures},\ }\href@noop {} {\bibfield  {journal} {\bibinfo  {journal} {Journal of Fluid Mechanics}\ }\textbf {\bibinfo {volume} {873}},\ \bibinfo {pages} {459} (\bibinfo {year} {2019})}\BibitemShut {NoStop}%
\bibitem [{\citenamefont {Alexakis}\ and\ \citenamefont {Biferale}(2018)}]{alexakis2018cascades}%
  \BibitemOpen
  \bibfield  {author} {\bibinfo {author} {\bibfnamefont {A.}~\bibnamefont {Alexakis}}\ and\ \bibinfo {author} {\bibfnamefont {L.}~\bibnamefont {Biferale}},\ }\bibfield  {title} {\bibinfo {title} {Cascades and transitions in turbulent flows},\ }\href@noop {} {\bibfield  {journal} {\bibinfo  {journal} {Physics Reports}\ }\textbf {\bibinfo {volume} {767}},\ \bibinfo {pages} {1} (\bibinfo {year} {2018})}\BibitemShut {NoStop}%
\bibitem [{\citenamefont {Verma}(2019)}]{verma2019energy}%
  \BibitemOpen
  \bibfield  {author} {\bibinfo {author} {\bibfnamefont {M.~K.}\ \bibnamefont {Verma}},\ }\href@noop {} {\emph {\bibinfo {title} {Energy transfers in fluid flows: multiscale and spectral perspectives}}}\ (\bibinfo  {publisher} {Cambridge University Press},\ \bibinfo {year} {2019})\BibitemShut {NoStop}%
\bibitem [{\citenamefont {Liu}\ \emph {et~al.}(2012)\citenamefont {Liu}, \citenamefont {Shelley},\ and\ \citenamefont {Zhang}}]{liu2012oscillations}%
  \BibitemOpen
  \bibfield  {author} {\bibinfo {author} {\bibfnamefont {B.}~\bibnamefont {Liu}}, \bibinfo {author} {\bibfnamefont {M.}~\bibnamefont {Shelley}},\ and\ \bibinfo {author} {\bibfnamefont {J.}~\bibnamefont {Zhang}},\ }\bibfield  {title} {\bibinfo {title} {Oscillations of a layer of viscoelastic fluid under steady forcing},\ }\href@noop {} {\bibfield  {journal} {\bibinfo  {journal} {Journal of Non-Newtonian Fluid Mechanics}\ }\textbf {\bibinfo {volume} {175-176}} (\bibinfo {year} {2012})}\BibitemShut {NoStop}%
\end{thebibliography}
\end{document}